\documentclass[superscriptaddress,reprint,amsmath,amssymb,aps,floatfix]{revtex4-1}

\usepackage{amsmath,gensymb,textcomp,bm,dcolumn,eurosym,array,tabu,multirow,nicefrac,color,subfigure,graphicx,upgreek}
\usepackage[colorlinks, linkcolor=blue, citecolor=blue, urlcolor=blue, breaklinks=true]{hyperref}

\usepackage{mathpazo,times} 

\begin{document}

\title{In-field entanglement distribution over a 96\,km-long submarine optical fibre}

\thanks{Correspondence and requests for materials should be addressed to S\"oren Wengerowsky and Rupert Ursin.}

\author{S\"oren Wengerowsky}
\email{soeren.wengerowsky@univie.ac.at}
\affiliation{Institute for Quantum Optics and Quantum Information - Vienna (IQOQI), Austrian Academy of Sciences, Vienna, Austria}
\affiliation{Vienna Center for Quantum Science and Technology (VCQ), Vienna, Austria}
\author{Siddarth Koduru Joshi}
\affiliation{Institute for Quantum Optics and Quantum Information - Vienna (IQOQI), Austrian Academy of Sciences, Vienna, Austria}
\affiliation{Vienna Center for Quantum Science and Technology (VCQ), Vienna, Austria}

\author{Fabian Steinlechner}
\affiliation{Institute for Quantum Optics and Quantum Information - Vienna (IQOQI), Austrian Academy of Sciences, Vienna, Austria}
\affiliation{Vienna Center for Quantum Science and Technology (VCQ), Vienna, Austria}

\author{Julien R. Zichi}
\affiliation{Department  of  Applied  Physics,  Royal  Institute  of  Technology  (KTH),  SE-106  91  Stockholm, Sweden}
\affiliation{Single Quantum B.V., 2628 CGH Delft, The Netherlands}

\author{Sergiy. M. Dobrovolskiy}
\affiliation{Single Quantum B.V., 2628 CGH Delft, The Netherlands}

\author{Ren\'{e} van der Molen}
\affiliation{Single Quantum B.V., 2628 CGH Delft, The Netherlands}

\author{Johannes W. N. Los}
\affiliation{Single Quantum B.V., 2628 CGH Delft, The Netherlands}

\author{Val Zwiller}
\affiliation{Department  of  Applied  Physics,  Royal  Institute  of  Technology  (KTH),  SE-106  91  Stockholm, Sweden}
\affiliation{Single Quantum B.V., 2628 CGH Delft, The Netherlands}

\author{Marijn A. M. Versteegh}
\affiliation{Department  of  Applied  Physics,  Royal  Institute  of  Technology  (KTH),  SE-106  91  Stockholm, Sweden}

\author{Alberto Mura}
\affiliation{I.N.Ri.M.---Istituto Nazionale  di Ricerca  Metrologica, Turin, Italy}

\author{Davide Calonico}
\affiliation{I.N.Ri.M.---Istituto Nazionale  di Ricerca  Metrologica, Turin, Italy}

\author{Massimo Inguscio}

\affiliation{LENS and Dipartimento di Fisica e Astronomia, Universit\`a di Firenze, 50019 Sesto Fiorentino, Italy}
\affiliation{CNR - Consiglio Nazionale delle Recerche, 00185 Rome, Italy}

\author{Hannes H\"ubel}
\affiliation{Security \& Communications Technologies; Center for Digital Safety \& Security, AIT Austrian Institute of Technology GmbH, Donau-City-Str. 1, 1220 Vienna, Austria}

\author{Anton Zeilinger}
\affiliation{Institute for Quantum Optics and Quantum Information - Vienna (IQOQI), Austrian Academy of Sciences, Vienna, Austria}
\affiliation{Quantum Optics, Quantum Nanophysics and Quantum Information, Faculty of Physics, University of Vienna, Boltzmanngasse 5, Vienna 1090, Austria}

\author{Andr\'{e} Xuereb}
\affiliation{Department of Physics, University of Malta, Msida MSD 2080, Malta}

\author{Rupert Ursin}
\email{rupert.ursin@oeaw.ac.at}
\affiliation{Institute for Quantum Optics and Quantum Information - Vienna (IQOQI), Austrian Academy of Sciences, Vienna, Austria}
\affiliation{Vienna Center for Quantum Science and Technology (VCQ), Vienna, Austria}

\begin{abstract}
Techniques for the distribution of quantum-secured cryptographic keys have reached a level of maturity allowing them to be implemented in all kinds of environments, away from any form of laboratory infrastructure. Here, we detail the distribution of entanglement between Malta and Sicily over a $96$\,km-long submarine telecommunications optical fibre cable. We used this standard telecommunications fibre as a quantum channel to distribute polarisation-entangled photons and were able to observe around $257$ photon pairs per second, with a polarisation visibility above $90$\%. 
Our experiment demonstrates the feasibility of using deployed submarine telecommunications optical fibres as long-distance quantum channels for polarisation-entangled photons. This opens up a plethora of possibilities for future experiments and technological applications using existing infrastructure.
\end{abstract}

\maketitle

Quantum cryptography promises a conceptual leap in information security. It proposes impenetrable physical-layer security for communication systems, based on the laws of quantum physics and not on assumptions on the limits of the computational power an adversary may or may not possess in the present or future. The past two decades have seen a flurry of work in quantum key distribution (QKD), and the underlying principles of the technique have been established by now ~\cite{bennett1992experimental,jennewein2000quantum,naik2000entangled,tittel2000quantum,muller1996quantum,Ursin2007}. Constant improvement has seen the experiments distribute cryptographic keys over ever-longer distances~\cite{Korzh2014,Yin2016} and increase the speed of key generation~\cite{zhang2009megabits, dixon2008gigahertz,eraerds2010_1gbps_singlefibre}. Intrinsically a point-to-point technology, QKD can however be combined with other ideas to extend quantum security to entire networks~\cite{Stucki2011,Sasaki2011,Peev2009,Xu2009,Elliott2005,courtland2016chinalink,wang2017cas_chinalink,wang2014fieldtest_wuhu_hefei}. Effort is also well underway to expand the reach of QKD systems, aiming at linking cities~\cite{wang2017cas_chinalink,muller1996quantum}; connecting continents with space-based links~\cite{liao2017satelliteqkd,Liao2018_intercontinentalqkd,takenaka2017satellite,gunthner2017quantum} has also been demonstrated, thus ushering in the possibility of a quantum-secure world-wide network. Aside from the possibility of truly secure long-distance communication, quantum networks and device-independent approaches in general are expected to have new applications, e.g., for randomness generation and distribution. 

Quantum entanglement holds the potential to form the basis of device-independent quantum-secure cryptography~\cite{masanes2011_device_indep,branciard2012_semi_device_indep} and is therefore especially interesting for applications. Amongst the many degrees of freedom which can be used to encode quantum informations, polarisation entangled qubits have the benefit that they are easy to measure and prepare with a high fidelity. Polarisation is an extraordinarily suitable carrier of quantum information over long distances, because there are no significant depolarisation effects over distances of the order of 100\,km, in both air and optical fibres. Changes in polarisation, which might happen, can in general be very well represented by unitary transformations and are therefore easily undone.

Polarisation-entangled qubits are used less frequently than time-bin qubits \cite{Marcikic2002} to distribute quantum information and quantum cryptographic keys over optical fibres, since the latter is conventionally regarded as more suitable~\cite{gisin2002,brodsky2011loss,antonelli2011sudden} for this purpose. 

\begin{figure*}[t]
 \centering
 \includegraphics[width=\linewidth]{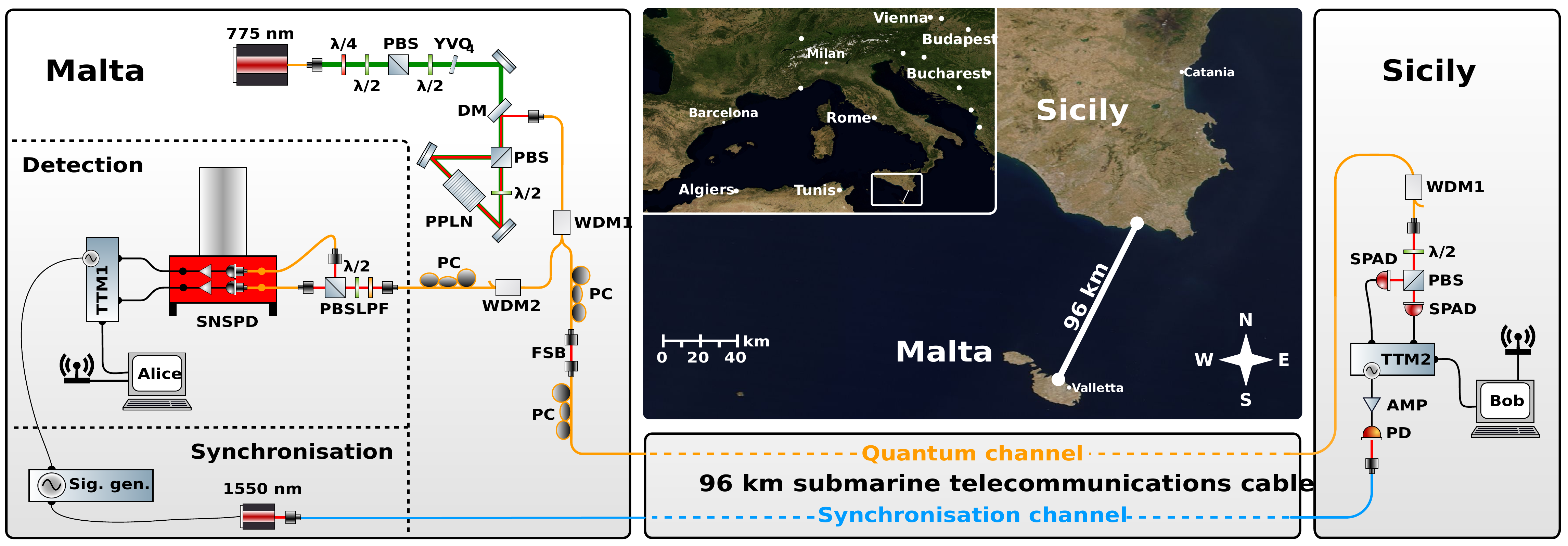}
 \caption{\label{fig_setup} Set-up and location of the experiment. The cable used here links the Mediterranean islands of Malta and Sicily. Photos courtesy of NASA Worldview~\cite{NASAMap1,NASAMap2}. A continuous-wave laser at 775\,nm bidirectionally pumped a PPLN crystal and created, via the process of spontaneous parametric down-conversion, photon pairs, which are entangled in polarisation due to the Sagnac geometry. Signal and idler photons are separated by frequency into two different fibres; one is detected locally in Malta in a polarisation analysis module consisting of a half-wave plate in front of a PBS and two superconducting nanowire detectors (SNSPD), and the other in Sicily after transmission through the 96\,km submarine telecommunications fibre. (Abbreviations: ---AMP:\ 50\,dB voltage amplifier; $\lambda/4$, $\lambda/2$:\ wave-plates; PBS:\ polarising beam-splitter, YVO$_4$:\ yttrium orthovanadate plate; DM:\ dichroic mirror; PPLN:\ MgO-doped periodically poled lithium niobate crystal (MgO:ppLN); WDM1: 100\,GHz band-pass filter (center wavelength: 1548.52\,nm), WDM2:\ 100\,GHz band-pass filter (center wavelength: 1551.72\,nm); PC:\ fibre polarisation controllers; LPF:\ 780\,nm long-pass filter; SNSPD:\  superconducting nanowire single photon detectors from Single Quantum; TTM1, TTM2:\ time-tagging modules; Sig.\ gen.:\ $10$\,MHz signal generator; SPAD:\ single-mode fibre coupled single-photon avalanche detectors; PD:\ fast InGaAs photodiode; FSB:\ free-space beam. Mirrors and fibre couplers not labelled, lenses omitted.)}
\end{figure*}

More generally, and despite the increasing level of maturity shown by some quantum technologies, it remains necessary to demonstrate the robustness of entanglement distribution required for its deployment in industrially relevant operational environments. While the distribution of entanglement via free-space~\cite{Ursin2007,yin2012_lake,marcikic2006freespacekurtsiefer} and satellite links~\cite{Yin2017a,ren2017_ground_to_satellite_teleportation} has seen tremendous advancement in the recent past, the vast majority of previous fibre-based experiments have been performed under idealised conditions such as a fibre coil inside a single laboratory~\cite{hubel2007high,Inagaki2013,Aktas2016,honjo2007_100km_timebin}. Notable exceptions include the distribution of time-bin entangled photons over 18\,km of telecommunications fiber \cite{salart2008testing} and  polarisation-entangled photon pairs over 1.45\,km~\cite{poppe2004practical} and 16\,km~\cite{treiber2009fully}, respectively.  Recently, quantum teleportation has been shown in deployed fibre networks, using time-bin encoding over 16\,km~\cite{valivarthi2016telep}  and polarisation-entangled photons~\cite{sun2016_chinese-teleportation-deployed-fiber}  over 30\,km.  Entanglement swapping using time-bin encoding has also been shown over 100\,km fibre, whereby the receivers were 12.5\,km apart~\cite{sun2017_entswapping_100km_timebin}.

Here, we distribute polarisation-entangled photons using a deployed submarine telecommunications optical fibre cable linking the Mediterranean islands of Malta and Sicily. The transmitter and receiver sites were separated by a distance of 93.4\,km and lacked any form of laboratory infrastructure.
We thus demonstrate conclusively, that polarisation-entanglement between two photons is well preserved over long distances in fibre in a real-world scenario and allows for a full implementation of QKD schemes over standard submarine telecommunications links.

\section{Results}

\textbf{Implementation.} In our experiment (Fig.~\ref{fig_setup}), we distributed entangled photons between the Mediterranean islands of Malta and Sicily. 
A source of polarisation-entangled photon pairs was located in Malta in the central data centre of one of the local telecommunication providers (Melita Ltd.), close to Fort Madliena. 
One photon from each pair was sent to a polarisation analysis and detection module located in Malta close to the source, consisting of a half-wave plate and a polarising beam-splitter with single photon detectors connected to the transmitted and reflected output ports. The other photon was sent to Sicily via a deployed $96$\,km-long submarine telecommunications optical fibre cable  which introduced an attenuation of $\sim22$\,dB. The link contains neither repeaters nor amplifiers of any sort and consists of several fibres which are all compliant with ITU-T G.655. Two of these fibres are used to transmit live classical communication in the C-band around 1550\,nm, and one dark fibre within the same cable represents the quantum channel in this experiment. At the other end of the fibre the second setup, consisting of an identical polarisation analysis and detection module, was set up inside a vehicle stationed outdoors close to the town of Pozzallo in Sicily; 
access to the optical fibre was obtained through a manhole. A separate optical fibre within the same cable was used to synchronise the two time-tagging modules by means of a laser operating at $1550$\,nm and modulated at $10$\,MHz by a signal generator. Two GPS clocks were used to help coordinate the experiment.

\textbf{Entangled-photon source and detection.}
The source was based on type-0 spontaneous parametric down-conversion in a 4\,cm-long Magnesium Oxide doped periodically poled,  Lithium Niobate (MgO:ppLN) crystal with a poling period of $19.2$\,$\upmu$m, temperature-stabilized at 85.4{\textdegree}C. The type-0 process converts, with a low probability, one pump photon at $775.075$\,nm from a continuous-wave laser to two photons, commonly named signal and idler photons, in the telecommunications C-band, having the same vertical polarisation as the pump photon. The MgO:ppLN crystal was bidirectionally pumped inside a Sagnac-type setup~\cite{Kim2005,lim2008}, thus creating the polarisation-entangled Bell state 
\begin{equation}\label{eq:state}
\lvert\Phi\rangle =\frac{1}{\sqrt{2}}(\lvert\text{V}_{\lambda_\text{s}}\text{V}_{\lambda_\text{i}}\rangle - \lvert\text{H}_{\lambda_\text{s}}\text{H}_{\lambda_\text{i}}\rangle),
\end{equation}
where we denote the signal (idler) wavelength by $\lambda_\text{s}$, ($\lambda_\text{i}$), and the polarisation degree of freedom by H (horizontal) or V (vertical). The joint spectrum of signal and idler has a full-width at half-maximum (FWHM) of approximately $60$\,nm.  Due to conservation of energy during the down-conversion process from a well-defined pump energy, polarisation-entangled photon pairs are found at equal spectral distance from the central frequency. We used approximately $0.6$\,nm FWHM band-pass filters to separate signal and idler photons at an equal spectral distance of the channels from the central wavelength of $1550.15$\,nm. ITU DWDM channel 36 ($\lambda_\text{s}=1548.52$\,nm) was chosen for the signal photons to be sent to Sicily while the idler photons in channel 32 ($\lambda_\text{i}=1551.72$\,nm) were detected locally in Malta. 

\begin{figure}[t]
 \centering
 \includegraphics[width=\columnwidth]{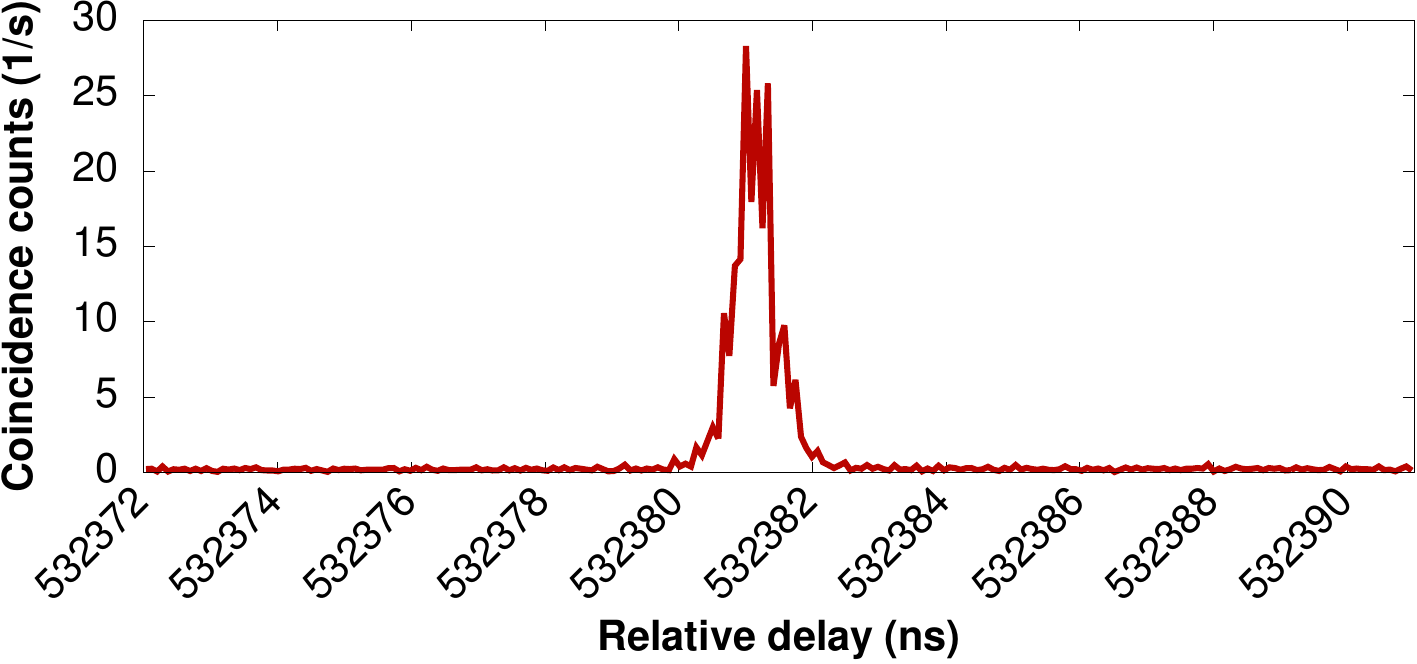}
 \caption{\label{fig_g2peak} The cross-correlation function between the time tags from Malta and Sicily shows a peak at a relative delay of approximately 532281\,ns, which corresponds to the length of the fibre, when we take into account the different latencies of the detection systems. The FWHM of approximately $0.7$\,ns is attributed to timing uncertainty of the SPADs in Sicily (approx. 400\,ps), the dispersion of the fibre link (approx. 500\,ps) and other effects related to the triggering and jitter in the time-tagging units (approx. 300\,ps).}
 \end{figure}

Locally in Malta, the visibility of the source was measured at approximately $98$\% in the diagonal--anti-diagonal (DA) polarisation basis and $97$\% in the horizontal--vertical (HV) basis. The local heralding efficiency was approximately $12$\%, measured on the superconducting nanowire single photon detector (SNSPD) system; roughly $28\,000$ pairs were detected per milliwatt of pump power. Two SNSPDs, necessary for handling the high countrates involved, were employed in the detection system in Malta and operated at an efficiency of approximately $54$\% and $59$\%, and with a dark-count rate of approximately $550$ and $470$ counts per second, respectively.

\begin{figure}[t]
 \centering
 \includegraphics[width=0.95\columnwidth]{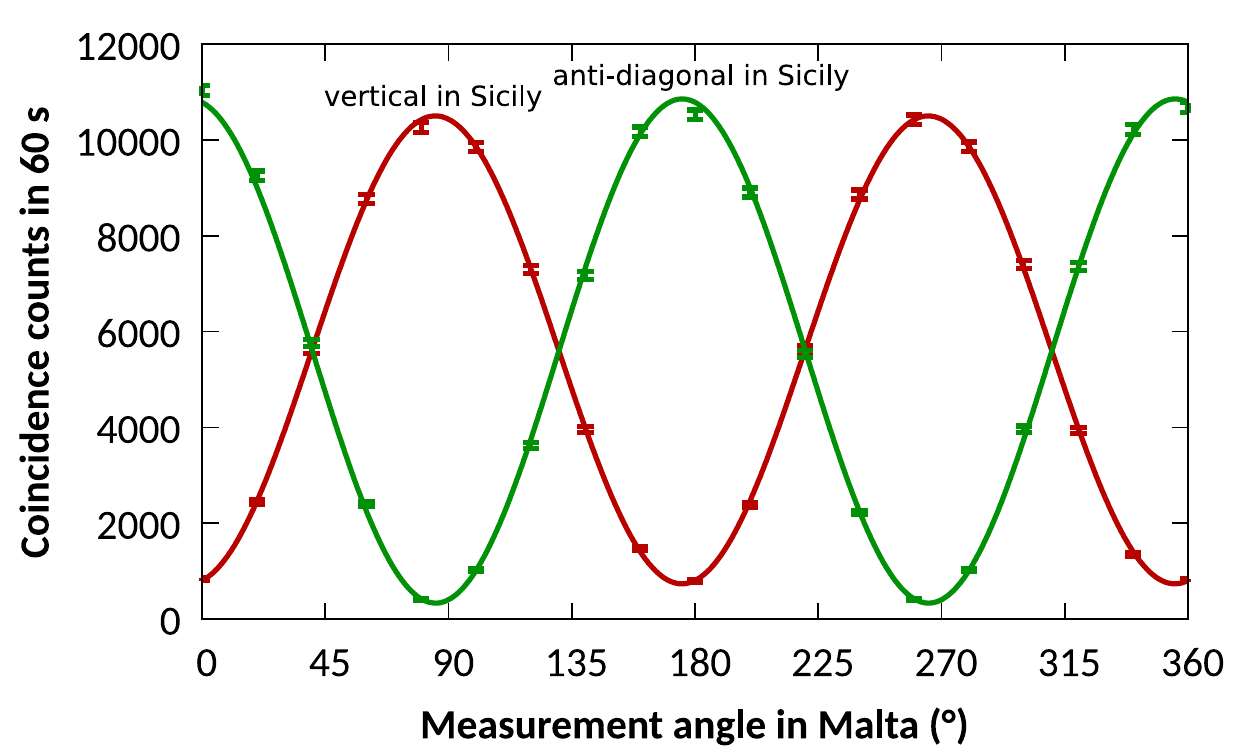}
 \caption{\label{fig_vis} Coincidence count rates for one detector pair and two different measurement angles in Sicily (vertical [red], and diagonal [green]) as a function of the  measurement angle for the analyser in Malta, $\phi_\text{M}$, starting from H [red] or D [green]. Poissonian statistics are assumed for the data as indicated by the error bars.}
\end{figure}

The detection system used in Sicily was more mobile and was in fact mounted in a vehicle that was moved to the location  and connected to the submarine fibre daily. This detection system used single-photon avalanche detectors (SPADs) which present very different characteristics to SNSPDs in terms of efficiency and dark counts. One detector had an efficiency of approximately $2$--$4$\% at a dead-time of $1$\,$\upmu$s and approximately $140$ dark counts per second, while the other operated at an efficiency of approximately $10$\% at a dead-time of $5$\,$\upmu$s with approximately $550$ dark counts per second.

\textbf{Entanglement characterisation.}
The photon arrival times were written to computer files locally and independently in Malta and Sicily. The two-photon coincidence events were identified by performing a cross-correlation between the time tags from Malta and Sicily, as shown in Fig.~\ref{fig_g2peak}. To quantify the quality of the entangled state after transmission through the submarine fibre, we performed a series of two-photon correlation measurements. In Sicily, the polarisation analyser was set to measure in either the H/V or D/A basis. The polarisation angle $\phi_\text{M}$ analysed in Malta was scanned from $0{\degree}$ to $360{\degree}$ in steps of $20{\degree}$. For each angle setting in Malta we accumulated data for a total of 60s. The best fit functions to the experimental data, two of which are shown in Fig.~\ref{fig_vis} exhibit a visibility of $86.8\%\pm0.8\%$ in the HV basis and $94.1\%\pm0.2\%$ in the DA basis. 

\begin{figure}[t]
 \centering
 \includegraphics[width=0.85\columnwidth]{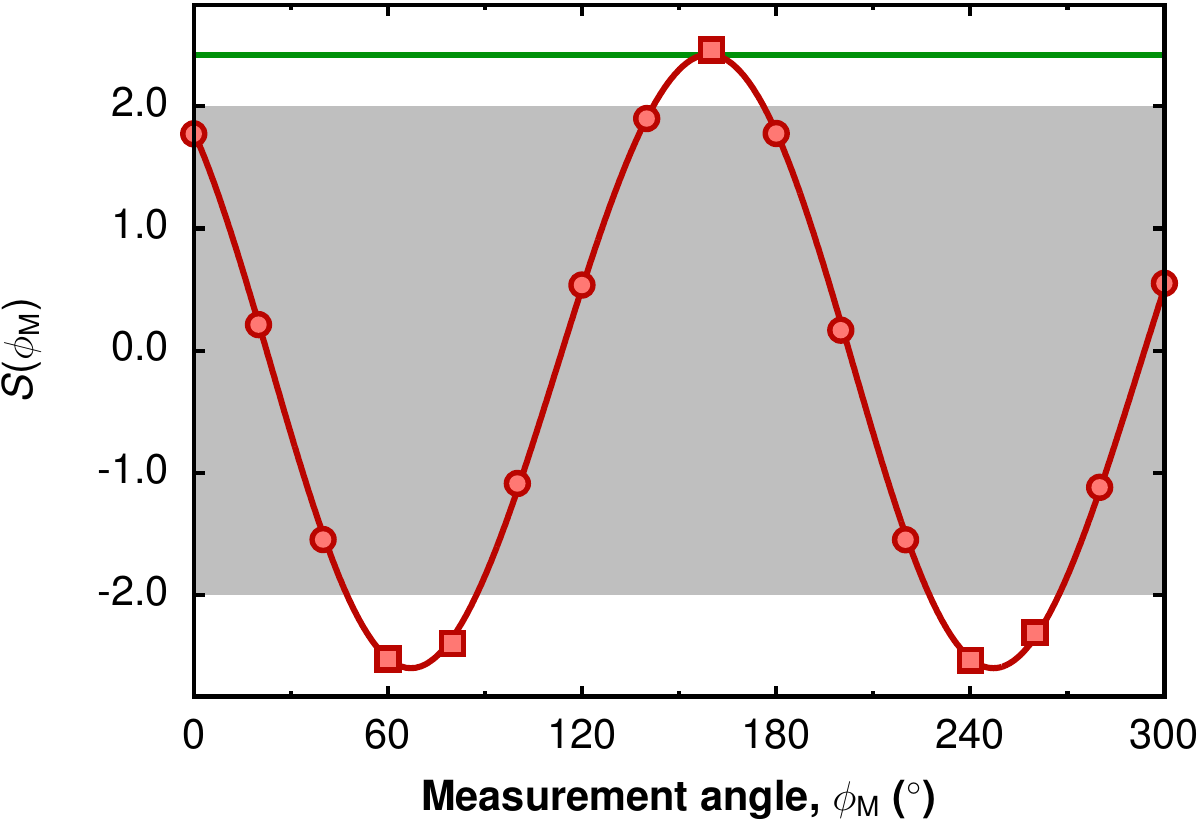}
 \caption{\label{fig_chsh} CHSH quantity $S(\phi_\text{M})$ as a function of the measurement angle for the analyser in Malta, $\phi_\text{M}$, which resembles the relative angle between the two mutually unbiased bases that were used in Malta and Sicily each; error bars are included but fit within the data markers; the standard deviation is $\leq0.014$ for all the points shown. Data outside the grey region (shown as squares) exclude local realistic theories. This function is computed using data similar to that shown in Fig.~\ref{fig_vis}. The solid red curve is obtained from a fit to the coincidence rates (as in Fig.~\ref{fig_vis}), not to the data shown here and yields a CHSH value of $2.534\pm 0.08$. The green horizontal line shows the CHSH value of $2.421\pm 0.008$ obtained in a measurement run at the fixed value of $\phi_\text{M}=157.5{\degree}$  (this includes a measurement at $22.5{\degree}$), i.e. the theoretically optimal angles.}
\end{figure}

\textbf{Bell test.}
To further quantify the quality of polarisation entanglement, we combined the results of the coincidence scans to yield the CHSH quantity $S(\phi_\text{M})$, which is bounded between $-2$ and $2$ for local realistic theories but may exceed these bounds up to an absolute value of $2\sqrt{2}$ in quantum mechanics and therefore serves as a strong witness for entanglement~\cite{CHSH}. To mitigate against systematic errors due to misalignment of the polarisation reference frames, we first used a best fit to the coincidence data (e.g. as shown in Fig.~\ref{fig_vis}) to compute $S(\phi_\text{M})$, as shown in Fig.~\ref{fig_chsh}.

 We observed the maximum Bell violation for a CHSH value of $-2.534\pm 0.08$, which corresponds to approximately 90\% of the Tsirelson bound~\cite{tsirelson1993quantum} in good agreement with the visibility of the two-photon coincidence data. Note that this value was obtained for $\phi_\text{M}=63.5{\degree}$, which corresponds to an offset of $4.0{\degree}$ from the theoretical optimum ($67.5{\degree}$). We ascribe this difference to a residual error in setting the zero point of our wave-plates and imperfect compensation of the fibres.

Finally, we directly measured the CHSH value with the analysers set to the theoretically optimal settings ($22.5{\degree}$ and $157.5{\degree}$) and (H/V and D/A), in Malta and Sicily, respectively. For each measurement setting, we accumulated data for a total of 600 seconds. This provided enough data to break down the data series into 39 blocks per measurement setting, and perform a statistical analysis of the data without having to rely on Poissonian count-rate statistics. The measured value for the CHSH quantity in this case is $2.421\pm0.008$ (as illustrated by the green horizontal line in Fig.~\ref{fig_chsh}), well beyond the bounds imposed by local realistic theories, consistent with the above value, and within the expected range.

The maximum coincidence rate we observed was $257\pm4$ counts per second, which corresponds to a secure key rate of approximately $30$ bits per second at the given visibility. The measured quantum bit error rate (QBER) was $5\%\pm0.5\%$, which is significantly better than the minimum requirement of 11\%~\cite{Ma2007}.

\section{Discussion}

We have successfully distributed polarisation-entangled photons over a $96$\,km-long submarine optical fibre link.  The deployed fibre is part of an existing telecommunications network linking the Mediterranean islands of Malta and Sicily. Our experiment marks the first international submarine quantum link.  It is also the longest distance distribution of entanglement in a deployed telecommunications network. We have demonstrated all the quantum prerequisites to be able to fully implement QKD. We verified the quality of entanglement by performing Bell tests and observed  a violation of the CHSH inequality at the level of $86$\% ($2.421$). This state was therefore observed with a fidelity high enough to enable secure quantum communication with QBER below $5.5$\% and estimated secure key rates of $30$ bits per second.  The link was stable within the statistical accuracy given by the assumed Poissonian statistics, and the QBER stayed constant for over two-and-a-half hours without active polarisation stabilisation. This is in accordance with results from other groups who investigated the changes of the polarisation state of aerial~\cite{waddy2005polarization} and buried~\cite{woodward2014long,ding2017polarization} fibres and found slow drifts on the scale of hours or days. Based on this we can conclusively prove that secure polarisation entanglement-based quantum communication is indeed possible over comparable deployed fibre links.

In previous field-trials of quantum cryptography, the opinion was voiced~\cite{gisin2002,gisin2007quantumcommunication} that polarisation states are not suitable for long-distance fibre-based QKD. In our opinion, this has changed due to enormous technological progress within the last decade. 
We believe that our data indicates that polarisation entanglement might actually be a very good choice, if not the preferred choice, for future entanglement-based quantum key distribution networks. 
A significiant advantage of polarisation entanglement is that it does not require interferometric measurements or interferometric state preparations, neither for sender or receiver, nor for the source. It also does not require external time synchronisation or any special techniques to overcome pulse broadening, as we demonstrate in the current paper. 

 Polarisation entanglement can also be used to seamlessly interface between free-space and fibre-based communication links. Finally, one can simply make use of the many quantum repeater and quantum networking schemes that have been demonstrated for polarisation entanglement, which can further extend the range of QKD systems and the number of clients they can reach. As an outlook, we note that using commercially available detectors with improved timing resolution, we could more than double the distance with respect to our present experiment. Our work thus opens up the possibility of using polarisation entanglement for truly global-scale fibre-based quantum communication.
 
 \section{Methods}
 \textbf{Fibre birefringence compensation.}
The $\lvert\Phi\rangle$ state was optimised locally in Malta by changing the polarisation of the pump beam and characterised using the local detection module and a polarisation analysis module that was inserted into the region denoted FSB in Fig.~\ref{fig_setup}. In order to ensure that the quantum state can be detected at the other end of the 96\,km fibre link, the polarisation rotation  of the quantum channel was neutralised by receiving alternately one of the two mutually unbiased polarisation states H and D from a laser, which was connected in the place of an SPAD in Sicily. The neutralization was done manually, using the signal of a polarimeter placed in the region FSB and manual fibre polarisation controllers.

 \textbf{Single-photon counting in Malta.}
The superconducting detectors used in Malta were fabricated from a newly-developed $9$\,nm-thick NbTiN superconducting film deposited by reactive co-sputtering at room temperature at the Swedish Royal Institute of Technology (KTH). The nanowires were patterned using electron-beam lithography and subsequent dry etching in collaboration with Single Quantum, and included further fabrication steps such as back-mirror integration and through-wafer etching for fibre alignment. We used a commercial cryostat (Single Quantum Eos), operating at $2.9$\,K and a current driver (Single Quantum Atlas) to operate the fibre-coupled SNSPDs. The efficiency of the detectors being dependent on the photon polarisation, a three-paddle fibre polarisation controller was used to optimise the detection efficiency. The SNSPD system operated continuously for two weeks in a data centre facility at an ambient temperature of about $30$\,{\degree}C without any degradation in performance.

\textbf{CHSH Measurements.} 
To compute the S-value, measurements from 4 basis settings were combined, while coincidence counts between all 4 detectors were used. The CHSH inequality reads: $-2\leq E(a_1,b_1)+E(a_1,b_2)+E(a_2,b_1)-E(a_2,b_2) \leq 2$, while $a_i$ with $i=1,2$ are the angles in Malta with $a_1-a_2 = 45\degree$ and $b_1-b_2=45\degree$ in Sicily. The correlation functions $E(a_i,b_i) $ are computed from the coincidence counts $C(a_i,b_j)$, measured at the angles $a_i,b_j$ as follows: $$E(a_i,b_j)=\frac{C(a_i,b_j)+C(a_{i \perp},b_{j \perp})-C(a_{i\perp},b_j)-C(a_i,b_{j,\perp})}{C(a_i,b_j)+C(a_{i \perp},b_{j \perp})+C(a_{i\perp},b_j)+C(a_i,b_{j,\perp})}$$
The symbol $\perp$ corresponds to the perpendicular angle, i.e. the second output of the polarising beam-splitter. 
The angle $\phi_{\text{M}}$ in Fig.~\ref{fig_chsh} can be understood as the relative angle between the measurement bases used in Malta and Sicily and is proportional to $a_i-b_i$.
\hspace{5cm}

\section{Acknowledgements}
We are deeply indebted to Simon Montanaro, \mbox{Roderick} Cassar, and Charles Peresso at Melita Ltd.\ for providing assistance and access to their network.  We thank Johannes Handsteiner for lending us two GPS clocks, Jesse Slim for programming a user interface for our motorised rotation stages, Evelyn Aracely Acu\~{n}a Ortega, Lukas Bulla, Matthias Fink,  Johannes Kofler,  Rosario Giordanella,  Thomas Scheidl, Leah Paula Vella and Ryan Vella for helpful discussions and technical assistance. 
We acknowledge the financial assistance of the University of Malta Research, Innovation \& Development Trust (RIDT). V.Z. acknowledges funding by the European Research Council under the grant agreement No. 307687 (NaQuOp) and the Swedish Research Council under grant agreement 638-2013-7152. Financial support was also provided by the Linnaeus Center in Advanced Optics and Photonics (Adopt). Financial support from the Austrian Research Promotion Agency (FFG) -Agentur f\"ur Luft- und Raumfahrt (FFG-ALR contract 844360 and FFG/ASAP:6238191 / 854022), the European Space Agency (ESA contract 4000112591/14/NL/US), the Austrian Science Fund (FWF) through (P24621-N27) and the START project (Y879-N27), as well as the Austrian Academy of Sciences is gratefully acknowledged.


\begin{thebibliography}{54}%
	\makeatletter
	\providecommand \@ifxundefined [1]{%
		\@ifx{#1\undefined}
	}%
	\providecommand \@ifnum [1]{%
		\ifnum #1\expandafter \@firstoftwo
		\else \expandafter \@secondoftwo
		\fi
	}%
	\providecommand \@ifx [1]{%
		\ifx #1\expandafter \@firstoftwo
		\else \expandafter \@secondoftwo
		\fi
	}%
	\providecommand \natexlab [1]{#1}%
	\providecommand \enquote  [1]{``#1''}%
	\providecommand \bibnamefont  [1]{#1}%
	\providecommand \bibfnamefont [1]{#1}%
	\providecommand \citenamefont [1]{#1}%
	\providecommand \href@noop [0]{\@secondoftwo}%
	\providecommand \href [0]{\begingroup \@sanitize@url \@href}%
	\providecommand \@href[1]{\@@startlink{#1}\@@href}%
	\providecommand \@@href[1]{\endgroup#1\@@endlink}%
	\providecommand \@sanitize@url [0]{\catcode `\\12\catcode `\$12\catcode
		`\&12\catcode `\#12\catcode `\^12\catcode `\_12\catcode `\%12\relax}%
	\providecommand \@@startlink[1]{}%
	\providecommand \@@endlink[0]{}%
	\providecommand \url  [0]{\begingroup\@sanitize@url \@url }%
	\providecommand \@url [1]{\endgroup\@href {#1}{\urlprefix }}%
	\providecommand \urlprefix  [0]{URL }%
	\providecommand \Eprint [0]{\href }%
	\providecommand \doibase [0]{http://dx.doi.org/}%
	\providecommand \selectlanguage [0]{\@gobble}%
	\providecommand \bibinfo  [0]{\@secondoftwo}%
	\providecommand \bibfield  [0]{\@secondoftwo}%
	\providecommand \translation [1]{[#1]}%
	\providecommand \BibitemOpen [0]{}%
	\providecommand \bibitemStop [0]{}%
	\providecommand \bibitemNoStop [0]{.\EOS\space}%
	\providecommand \EOS [0]{\spacefactor3000\relax}%
	\providecommand \BibitemShut  [1]{\csname bibitem#1\endcsname}%
	\let\auto@bib@innerbib\@empty
	\bibitem [{\citenamefont {Bennett}\ \emph {et~al.}(1992)\citenamefont
		{Bennett}, \citenamefont {Bessette}, \citenamefont {Brassard}, \citenamefont
		{Salvail},\ and\ \citenamefont {Smolin}}]{bennett1992experimental}%
	\BibitemOpen
	\bibfield  {author} {\bibinfo {author} {\bibfnamefont {C.}~\bibnamefont
			{Bennett}}, \bibinfo {author} {\bibfnamefont {F.}~\bibnamefont {Bessette}},
		\bibinfo {author} {\bibfnamefont {G.}~\bibnamefont {Brassard}}, \bibinfo
		{author} {\bibfnamefont {L.}~\bibnamefont {Salvail}}, \ and\ \bibinfo
		{author} {\bibfnamefont {J.}~\bibnamefont {Smolin}},\ }\href {\doibase
		10.1007/BF00191318} {\bibfield  {journal} {\bibinfo  {journal} {Journal of
				Cryptology}\ }\textbf {\bibinfo {volume} {5}},\ \bibinfo {pages} {3}
		(\bibinfo {year} {1992})}\BibitemShut {NoStop}%
	\bibitem [{\citenamefont {Jennewein}\ \emph {et~al.}(2000)\citenamefont
		{Jennewein}, \citenamefont {Simon}, \citenamefont {Weihs}, \citenamefont
		{Weinfurter},\ and\ \citenamefont {Zeilinger}}]{jennewein2000quantum}%
	\BibitemOpen
	\bibfield  {author} {\bibinfo {author} {\bibfnamefont {T.}~\bibnamefont
			{Jennewein}}, \bibinfo {author} {\bibfnamefont {C.}~\bibnamefont {Simon}},
		\bibinfo {author} {\bibfnamefont {G.}~\bibnamefont {Weihs}}, \bibinfo
		{author} {\bibfnamefont {H.}~\bibnamefont {Weinfurter}}, \ and\ \bibinfo
		{author} {\bibfnamefont {A.}~\bibnamefont {Zeilinger}},\ }\href {\doibase
		10.1103/PhysRevLett.84.4729} {\bibfield  {journal} {\bibinfo  {journal}
			{Physical Review Letters}\ }\textbf {\bibinfo {volume} {84}},\ \bibinfo
		{pages} {4729} (\bibinfo {year} {2000})}\BibitemShut {NoStop}%
	\bibitem [{\citenamefont {Naik}\ \emph {et~al.}(2000)\citenamefont {Naik},
		\citenamefont {Peterson}, \citenamefont {White}, \citenamefont {Berglund},\
		and\ \citenamefont {Kwiat}}]{naik2000entangled}%
	\BibitemOpen
	\bibfield  {author} {\bibinfo {author} {\bibfnamefont {D.~S.}\ \bibnamefont
			{Naik}}, \bibinfo {author} {\bibfnamefont {C.~G.}\ \bibnamefont {Peterson}},
		\bibinfo {author} {\bibfnamefont {A.~G.}\ \bibnamefont {White}}, \bibinfo
		{author} {\bibfnamefont {A.~J.}\ \bibnamefont {Berglund}}, \ and\ \bibinfo
		{author} {\bibfnamefont {P.~G.}\ \bibnamefont {Kwiat}},\ }\href {\doibase
		10.1103/PhysRevLett.84.4733} {\bibfield  {journal} {\bibinfo  {journal}
			{Physical Review Letters}\ }\textbf {\bibinfo {volume} {84}},\ \bibinfo
		{pages} {4733} (\bibinfo {year} {2000})}\BibitemShut {NoStop}%
	\bibitem [{\citenamefont {Tittel}\ \emph {et~al.}(2000)\citenamefont {Tittel},
		\citenamefont {Brendel}, \citenamefont {Zbinden},\ and\ \citenamefont
		{Gisin}}]{tittel2000quantum}%
	\BibitemOpen
	\bibfield  {author} {\bibinfo {author} {\bibfnamefont {W.}~\bibnamefont
			{Tittel}}, \bibinfo {author} {\bibfnamefont {J.}~\bibnamefont {Brendel}},
		\bibinfo {author} {\bibfnamefont {H.}~\bibnamefont {Zbinden}}, \ and\
		\bibinfo {author} {\bibfnamefont {N.}~\bibnamefont {Gisin}},\ }\href
	{\doibase 10.1103/PhysRevLett.84.4737} {\bibfield  {journal} {\bibinfo
			{journal} {Physical Review Letters}\ }\textbf {\bibinfo {volume} {84}},\
		\bibinfo {pages} {4737} (\bibinfo {year} {2000})}\BibitemShut {NoStop}%
	\bibitem [{\citenamefont {Muller}\ \emph {et~al.}(1996)\citenamefont {Muller},
		\citenamefont {Zbinden},\ and\ \citenamefont {Gisin}}]{muller1996quantum}%
	\BibitemOpen
	\bibfield  {author} {\bibinfo {author} {\bibfnamefont {A.}~\bibnamefont
			{Muller}}, \bibinfo {author} {\bibfnamefont {H.}~\bibnamefont {Zbinden}}, \
		and\ \bibinfo {author} {\bibfnamefont {N.}~\bibnamefont {Gisin}},\ }\href
	{\doibase 10.1209/epl/i1996-00343-4} {\bibfield  {journal} {\bibinfo
			{journal} {Europhysics Letters (EPL)}\ }\textbf {\bibinfo {volume} {33}},\
		\bibinfo {pages} {335} (\bibinfo {year} {1996})}\BibitemShut {NoStop}%
	\bibitem [{\citenamefont {Nagali}\ \emph {et~al.}(2012)\citenamefont {Nagali},
		\citenamefont {D'Ambrosio}, \citenamefont {Sciarrino},\ and\ \citenamefont
		{Cabello}}]{Ursin2007}%
	\BibitemOpen
	\bibfield  {author} {\bibinfo {author} {\bibfnamefont {E.}~\bibnamefont
			{Nagali}}, \bibinfo {author} {\bibfnamefont {V.}~\bibnamefont {D'Ambrosio}},
		\bibinfo {author} {\bibfnamefont {F.}~\bibnamefont {Sciarrino}}, \ and\
		\bibinfo {author} {\bibfnamefont {A.}~\bibnamefont {Cabello}},\ }\href
	{\doibase 10.1103/PhysRevLett.108.090501} {\bibfield  {journal} {\bibinfo
			{journal} {Nature Physics}\ }\textbf {\bibinfo {volume} {3}},\ \bibinfo
		{pages} {481} (\bibinfo {year} {2012})}\BibitemShut {NoStop}%
	\bibitem [{\citenamefont {Korzh}\ \emph {et~al.}(2015)\citenamefont {Korzh},
		\citenamefont {Lim}, \citenamefont {Houlmann}, \citenamefont {Gisin},
		\citenamefont {Li}, \citenamefont {Nolan}, \citenamefont {Sanguinetti},
		\citenamefont {Thew},\ and\ \citenamefont {Zbinden}}]{Korzh2014}%
	\BibitemOpen
	\bibfield  {author} {\bibinfo {author} {\bibfnamefont {B.}~\bibnamefont
			{Korzh}}, \bibinfo {author} {\bibfnamefont {C.~C.~W.}\ \bibnamefont {Lim}},
		\bibinfo {author} {\bibfnamefont {R.}~\bibnamefont {Houlmann}}, \bibinfo
		{author} {\bibfnamefont {N.}~\bibnamefont {Gisin}}, \bibinfo {author}
		{\bibfnamefont {M.~J.}\ \bibnamefont {Li}}, \bibinfo {author} {\bibfnamefont
			{D.}~\bibnamefont {Nolan}}, \bibinfo {author} {\bibfnamefont
			{B.}~\bibnamefont {Sanguinetti}}, \bibinfo {author} {\bibfnamefont
			{R.}~\bibnamefont {Thew}}, \ and\ \bibinfo {author} {\bibfnamefont
			{H.}~\bibnamefont {Zbinden}},\ }\href {\doibase 10.1038/nphoton.2014.327}
	{\bibfield  {journal} {\bibinfo  {journal} {Nature Photonics}\ }\textbf
		{\bibinfo {volume} {9}},\ \bibinfo {pages} {163} (\bibinfo {year}
		{2015})}\BibitemShut {NoStop}%
	\bibitem [{\citenamefont {Yin}\ \emph {et~al.}(2016)\citenamefont {Yin},
		\citenamefont {Chen}, \citenamefont {Yu}, \citenamefont {Liu}, \citenamefont
		{You}, \citenamefont {Zhou}, \citenamefont {Chen}, \citenamefont {Mao},
		\citenamefont {Huang}, \citenamefont {Zhang}, \citenamefont {Chen},
		\citenamefont {Li}, \citenamefont {Nolan}, \citenamefont {Zhou},
		\citenamefont {Jiang}, \citenamefont {Wang}, \citenamefont {Zhang},
		\citenamefont {Wang},\ and\ \citenamefont {Pan}}]{Yin2016}%
	\BibitemOpen
	\bibfield  {author} {\bibinfo {author} {\bibfnamefont {H.~L.}\ \bibnamefont
			{Yin}}, \bibinfo {author} {\bibfnamefont {T.~Y.}\ \bibnamefont {Chen}},
		\bibinfo {author} {\bibfnamefont {Z.~W.}\ \bibnamefont {Yu}}, \bibinfo
		{author} {\bibfnamefont {H.}~\bibnamefont {Liu}}, \bibinfo {author}
		{\bibfnamefont {L.~X.}\ \bibnamefont {You}}, \bibinfo {author} {\bibfnamefont
			{Y.~H.}\ \bibnamefont {Zhou}}, \bibinfo {author} {\bibfnamefont {S.~J.}\
			\bibnamefont {Chen}}, \bibinfo {author} {\bibfnamefont {Y.}~\bibnamefont
			{Mao}}, \bibinfo {author} {\bibfnamefont {M.~Q.}\ \bibnamefont {Huang}},
		\bibinfo {author} {\bibfnamefont {W.~J.}\ \bibnamefont {Zhang}}, \bibinfo
		{author} {\bibfnamefont {H.}~\bibnamefont {Chen}}, \bibinfo {author}
		{\bibfnamefont {M.~J.}\ \bibnamefont {Li}}, \bibinfo {author} {\bibfnamefont
			{D.}~\bibnamefont {Nolan}}, \bibinfo {author} {\bibfnamefont
			{F.}~\bibnamefont {Zhou}}, \bibinfo {author} {\bibfnamefont {X.}~\bibnamefont
			{Jiang}}, \bibinfo {author} {\bibfnamefont {Z.}~\bibnamefont {Wang}},
		\bibinfo {author} {\bibfnamefont {Q.}~\bibnamefont {Zhang}}, \bibinfo
		{author} {\bibfnamefont {X.~B.}\ \bibnamefont {Wang}}, \ and\ \bibinfo
		{author} {\bibfnamefont {J.~W.}\ \bibnamefont {Pan}},\ }\href {\doibase
		10.1103/PhysRevLett.117.190501} {\bibfield  {journal} {\bibinfo  {journal}
			{Physical Review Letters}\ }\textbf {\bibinfo {volume} {117}},\ \bibinfo
		{pages} {190501} (\bibinfo {year} {2016})}\BibitemShut {NoStop}%
	\bibitem [{\citenamefont {Zhang}\ \emph {et~al.}(2008)\citenamefont {Zhang},
		\citenamefont {Takesue}, \citenamefont {Honjo}, \citenamefont {Wen},
		\citenamefont {Hirohata}, \citenamefont {Suyama}, \citenamefont {Takiguchi},
		\citenamefont {Kamada}, \citenamefont {Tokura}, \citenamefont {Tadanaga},
		\citenamefont {Nishida}, \citenamefont {Asobe},\ and\ \citenamefont
		{Yamamoto}}]{zhang2009megabits}%
	\BibitemOpen
	\bibfield  {author} {\bibinfo {author} {\bibfnamefont {Q.}~\bibnamefont
			{Zhang}}, \bibinfo {author} {\bibfnamefont {H.}~\bibnamefont {Takesue}},
		\bibinfo {author} {\bibfnamefont {T.}~\bibnamefont {Honjo}}, \bibinfo
		{author} {\bibfnamefont {K.}~\bibnamefont {Wen}}, \bibinfo {author}
		{\bibfnamefont {T.}~\bibnamefont {Hirohata}}, \bibinfo {author}
		{\bibfnamefont {M.}~\bibnamefont {Suyama}}, \bibinfo {author} {\bibfnamefont
			{Y.}~\bibnamefont {Takiguchi}}, \bibinfo {author} {\bibfnamefont
			{H.}~\bibnamefont {Kamada}}, \bibinfo {author} {\bibfnamefont
			{Y.}~\bibnamefont {Tokura}}, \bibinfo {author} {\bibfnamefont
			{O.}~\bibnamefont {Tadanaga}}, \bibinfo {author} {\bibfnamefont
			{Y.}~\bibnamefont {Nishida}}, \bibinfo {author} {\bibfnamefont
			{M.}~\bibnamefont {Asobe}}, \ and\ \bibinfo {author} {\bibfnamefont
			{Y.}~\bibnamefont {Yamamoto}},\ }\href {\doibase
		10.1088/1367-2630/11/4/045010} {\bibfield  {journal} {\bibinfo  {journal}
			{New Journal of Physics}\ }\textbf {\bibinfo {volume} {11}},\ \bibinfo
		{pages} {045010} (\bibinfo {year} {2008})}\BibitemShut {NoStop}%
	\bibitem [{\citenamefont {Dixon}\ \emph {et~al.}(2008)\citenamefont {Dixon},
		\citenamefont {Yuan}, \citenamefont {Dynes}, \citenamefont {Sharpe},\ and\
		\citenamefont {Shields}}]{dixon2008gigahertz}%
	\BibitemOpen
	\bibfield  {author} {\bibinfo {author} {\bibfnamefont {A.~R.}\ \bibnamefont
			{Dixon}}, \bibinfo {author} {\bibfnamefont {Z.~L.}\ \bibnamefont {Yuan}},
		\bibinfo {author} {\bibfnamefont {J.~F.}\ \bibnamefont {Dynes}}, \bibinfo
		{author} {\bibfnamefont {A.~W.}\ \bibnamefont {Sharpe}}, \ and\ \bibinfo
		{author} {\bibfnamefont {A.~J.}\ \bibnamefont {Shields}},\ }\href {\doibase
		10.1364/OE.16.018790} {\bibfield  {journal} {\bibinfo  {journal} {Optics
				Express}\ }\textbf {\bibinfo {volume} {16}},\ \bibinfo {pages} {18790}
		(\bibinfo {year} {2008})}\BibitemShut {NoStop}%
	\bibitem [{\citenamefont {Eraerds}\ \emph {et~al.}(2010)\citenamefont
		{Eraerds}, \citenamefont {Walenta}, \citenamefont {Legr{\'{e}}},
		\citenamefont {Gisin},\ and\ \citenamefont
		{Zbinden}}]{eraerds2010_1gbps_singlefibre}%
	\BibitemOpen
	\bibfield  {author} {\bibinfo {author} {\bibfnamefont {P.}~\bibnamefont
			{Eraerds}}, \bibinfo {author} {\bibfnamefont {N.}~\bibnamefont {Walenta}},
		\bibinfo {author} {\bibfnamefont {M.}~\bibnamefont {Legr{\'{e}}}}, \bibinfo
		{author} {\bibfnamefont {N.}~\bibnamefont {Gisin}}, \ and\ \bibinfo {author}
		{\bibfnamefont {H.}~\bibnamefont {Zbinden}},\ }\href {\doibase
		10.1088/1367-2630/12/6/063027} {\bibfield  {journal} {\bibinfo  {journal}
			{New Journal of Physics}\ }\textbf {\bibinfo {volume} {12}},\ \bibinfo
		{pages} {063027} (\bibinfo {year} {2010})}\BibitemShut {NoStop}%
	\bibitem [{\citenamefont {Stucki}\ \emph {et~al.}(2012)\citenamefont {Stucki},
		\citenamefont {Legre}, \citenamefont {Buntschu}, \citenamefont {Clausen},
		\citenamefont {Felber}, \citenamefont {Gisin}, \citenamefont {Henzen},
		\citenamefont {Junod}, \citenamefont {Litzistorf}, \citenamefont {Monbaron},
		\citenamefont {Monat}, \citenamefont {Page}, \citenamefont {Perroud},
		\citenamefont {Ribordy}, \citenamefont {Rochas}, \citenamefont {Robyr},
		\citenamefont {Tavares}, \citenamefont {Thew}, \citenamefont {Trinkler},
		\citenamefont {Ventura}, \citenamefont {Voirol}, \citenamefont {Walenta},\
		and\ \citenamefont {Zbinden}}]{Stucki2011}%
	\BibitemOpen
	\bibfield  {author} {\bibinfo {author} {\bibfnamefont {D.}~\bibnamefont
			{Stucki}}, \bibinfo {author} {\bibfnamefont {M.}~\bibnamefont {Legre}},
		\bibinfo {author} {\bibfnamefont {F.}~\bibnamefont {Buntschu}}, \bibinfo
		{author} {\bibfnamefont {B.}~\bibnamefont {Clausen}}, \bibinfo {author}
		{\bibfnamefont {N.}~\bibnamefont {Felber}}, \bibinfo {author} {\bibfnamefont
			{N.}~\bibnamefont {Gisin}}, \bibinfo {author} {\bibfnamefont
			{L.}~\bibnamefont {Henzen}}, \bibinfo {author} {\bibfnamefont
			{P.}~\bibnamefont {Junod}}, \bibinfo {author} {\bibfnamefont
			{G.}~\bibnamefont {Litzistorf}}, \bibinfo {author} {\bibfnamefont
			{P.}~\bibnamefont {Monbaron}}, \bibinfo {author} {\bibfnamefont
			{L.}~\bibnamefont {Monat}}, \bibinfo {author} {\bibfnamefont {J.~B.}\
			\bibnamefont {Page}}, \bibinfo {author} {\bibfnamefont {D.}~\bibnamefont
			{Perroud}}, \bibinfo {author} {\bibfnamefont {G.}~\bibnamefont {Ribordy}},
		\bibinfo {author} {\bibfnamefont {A.}~\bibnamefont {Rochas}}, \bibinfo
		{author} {\bibfnamefont {S.}~\bibnamefont {Robyr}}, \bibinfo {author}
		{\bibfnamefont {J.}~\bibnamefont {Tavares}}, \bibinfo {author} {\bibfnamefont
			{R.}~\bibnamefont {Thew}}, \bibinfo {author} {\bibfnamefont {P.}~\bibnamefont
			{Trinkler}}, \bibinfo {author} {\bibfnamefont {S.}~\bibnamefont {Ventura}},
		\bibinfo {author} {\bibfnamefont {R.}~\bibnamefont {Voirol}}, \bibinfo
		{author} {\bibfnamefont {N.}~\bibnamefont {Walenta}}, \ and\ \bibinfo
		{author} {\bibfnamefont {H.}~\bibnamefont {Zbinden}},\ }\href {\doibase
		10.1088/1367-2630/13/12/123001} {\bibfield  {journal} {\bibinfo  {journal}
			{New Journal of Physics}\ }\textbf {\bibinfo {volume} {13}},\ \bibinfo
		{pages} {123001} (\bibinfo {year} {2012})}\BibitemShut {NoStop}%
	\bibitem [{\citenamefont {Sasaki}\ \emph {et~al.}(2011)\citenamefont {Sasaki},
		\citenamefont {Fujiwara}, \citenamefont {Ishizuka}, \citenamefont {Klaus},
		\citenamefont {Wakui}, \citenamefont {Takeoka}, \citenamefont {Tanaka},
		\citenamefont {Yoshino}, \citenamefont {Nambu}, \citenamefont {Takahashi},
		\citenamefont {Tajima}, \citenamefont {Tomita}, \citenamefont {Domeki},
		\citenamefont {Hasegawa}, \citenamefont {Sakai}, \citenamefont {Kobayashi},
		\citenamefont {Asai}, \citenamefont {Shimizu}, \citenamefont {Tokura},
		\citenamefont {Tsurumaru}, \citenamefont {Matsui}, \citenamefont {Honjo},
		\citenamefont {Tamaki}, \citenamefont {Takesue}, \citenamefont {Tokura},
		\citenamefont {Dynes}, \citenamefont {Dixon}, \citenamefont {Sharpe},
		\citenamefont {Yuan}, \citenamefont {Shields}, \citenamefont {Uchikoga},
		\citenamefont {Legre}, \citenamefont {Robyr}, \citenamefont {Trinkler},
		\citenamefont {Monat}, \citenamefont {Page}, \citenamefont {Ribordy},
		\citenamefont {Poppe}, \citenamefont {Allacher}, \citenamefont {Maurhart},
		\citenamefont {Langer}, \citenamefont {Peev},\ and\ \citenamefont
		{Zeilinger}}]{Sasaki2011}%
	\BibitemOpen
	\bibfield  {author} {\bibinfo {author} {\bibfnamefont {M.}~\bibnamefont
			{Sasaki}}, \bibinfo {author} {\bibfnamefont {M.}~\bibnamefont {Fujiwara}},
		\bibinfo {author} {\bibfnamefont {H.}~\bibnamefont {Ishizuka}}, \bibinfo
		{author} {\bibfnamefont {W.}~\bibnamefont {Klaus}}, \bibinfo {author}
		{\bibfnamefont {K.}~\bibnamefont {Wakui}}, \bibinfo {author} {\bibfnamefont
			{M.}~\bibnamefont {Takeoka}}, \bibinfo {author} {\bibfnamefont
			{A.}~\bibnamefont {Tanaka}}, \bibinfo {author} {\bibfnamefont
			{K.}~\bibnamefont {Yoshino}}, \bibinfo {author} {\bibfnamefont
			{Y.}~\bibnamefont {Nambu}}, \bibinfo {author} {\bibfnamefont
			{S.}~\bibnamefont {Takahashi}}, \bibinfo {author} {\bibfnamefont
			{A.}~\bibnamefont {Tajima}}, \bibinfo {author} {\bibfnamefont
			{A.}~\bibnamefont {Tomita}}, \bibinfo {author} {\bibfnamefont
			{T.}~\bibnamefont {Domeki}}, \bibinfo {author} {\bibfnamefont
			{T.}~\bibnamefont {Hasegawa}}, \bibinfo {author} {\bibfnamefont
			{Y.}~\bibnamefont {Sakai}}, \bibinfo {author} {\bibfnamefont
			{H.}~\bibnamefont {Kobayashi}}, \bibinfo {author} {\bibfnamefont
			{T.}~\bibnamefont {Asai}}, \bibinfo {author} {\bibfnamefont {K.}~\bibnamefont
			{Shimizu}}, \bibinfo {author} {\bibfnamefont {T.}~\bibnamefont {Tokura}},
		\bibinfo {author} {\bibfnamefont {T.}~\bibnamefont {Tsurumaru}}, \bibinfo
		{author} {\bibfnamefont {M.}~\bibnamefont {Matsui}}, \bibinfo {author}
		{\bibfnamefont {T.}~\bibnamefont {Honjo}}, \bibinfo {author} {\bibfnamefont
			{K.}~\bibnamefont {Tamaki}}, \bibinfo {author} {\bibfnamefont
			{H.}~\bibnamefont {Takesue}}, \bibinfo {author} {\bibfnamefont
			{Y.}~\bibnamefont {Tokura}}, \bibinfo {author} {\bibfnamefont {J.~F.}\
			\bibnamefont {Dynes}}, \bibinfo {author} {\bibfnamefont {A.~R.}\ \bibnamefont
			{Dixon}}, \bibinfo {author} {\bibfnamefont {A.~W.}\ \bibnamefont {Sharpe}},
		\bibinfo {author} {\bibfnamefont {Z.~L.}\ \bibnamefont {Yuan}}, \bibinfo
		{author} {\bibfnamefont {A.~J.}\ \bibnamefont {Shields}}, \bibinfo {author}
		{\bibfnamefont {S.}~\bibnamefont {Uchikoga}}, \bibinfo {author}
		{\bibfnamefont {M.}~\bibnamefont {Legre}}, \bibinfo {author} {\bibfnamefont
			{S.}~\bibnamefont {Robyr}}, \bibinfo {author} {\bibfnamefont
			{P.}~\bibnamefont {Trinkler}}, \bibinfo {author} {\bibfnamefont
			{L.}~\bibnamefont {Monat}}, \bibinfo {author} {\bibfnamefont {J.~B.}\
			\bibnamefont {Page}}, \bibinfo {author} {\bibfnamefont {G.}~\bibnamefont
			{Ribordy}}, \bibinfo {author} {\bibfnamefont {A.}~\bibnamefont {Poppe}},
		\bibinfo {author} {\bibfnamefont {A.}~\bibnamefont {Allacher}}, \bibinfo
		{author} {\bibfnamefont {O.}~\bibnamefont {Maurhart}}, \bibinfo {author}
		{\bibfnamefont {T.}~\bibnamefont {Langer}}, \bibinfo {author} {\bibfnamefont
			{M.}~\bibnamefont {Peev}}, \ and\ \bibinfo {author} {\bibfnamefont
			{A.}~\bibnamefont {Zeilinger}},\ }\href {\doibase 10.1364/OE.19.010387}
	{\bibfield  {journal} {\bibinfo  {journal} {Optics Express}\ }\textbf
		{\bibinfo {volume} {19}},\ \bibinfo {pages} {10387} (\bibinfo {year}
		{2011})}\BibitemShut {NoStop}%
	\bibitem [{\citenamefont {Peev}\ \emph {et~al.}(2009)\citenamefont {Peev},
		\citenamefont {Pacher}, \citenamefont {All{\'{e}}aume}, \citenamefont
		{Barreiro}, \citenamefont {Bouda}, \citenamefont {Boxleitner}, \citenamefont
		{Debuisschert}, \citenamefont {Diamanti}, \citenamefont {Dianati},
		\citenamefont {Dynes}, \citenamefont {Fasel}, \citenamefont {Fossier},
		\citenamefont {F{\"{u}}rst}, \citenamefont {Gautier}, \citenamefont {Gay},
		\citenamefont {Gisin}, \citenamefont {Grangier}, \citenamefont {Happe},
		\citenamefont {Hasani}, \citenamefont {Hentschel}, \citenamefont
		{H{\"{u}}bel}, \citenamefont {Humer}, \citenamefont {L{\"{a}}nger},
		\citenamefont {Legr{\'{e}}}, \citenamefont {Lieger}, \citenamefont
		{Lodewyck}, \citenamefont {Lor{\"{u}}nser}, \citenamefont {L{\"{u}}tkenhaus},
		\citenamefont {Marhold}, \citenamefont {Matyus}, \citenamefont {Maurhart},
		\citenamefont {Monat}, \citenamefont {Nauerth}, \citenamefont {Page},
		\citenamefont {Poppe}, \citenamefont {Querasser}, \citenamefont {Ribordy},
		\citenamefont {Robyr}, \citenamefont {Salvail}, \citenamefont {Sharpe},
		\citenamefont {Shields}, \citenamefont {Stucki}, \citenamefont {Suda},
		\citenamefont {Tamas}, \citenamefont {Themel}, \citenamefont {Thew},
		\citenamefont {Thoma}, \citenamefont {Treiber}, \citenamefont {Trinkler},
		\citenamefont {Tualle-Brouri}, \citenamefont {Vannel}, \citenamefont
		{Walenta}, \citenamefont {Weier}, \citenamefont {Weinfurter}, \citenamefont
		{Wimberger}, \citenamefont {Yuan}, \citenamefont {Zbinden},\ and\
		\citenamefont {Zeilinger}}]{Peev2009}%
	\BibitemOpen
	\bibfield  {author} {\bibinfo {author} {\bibfnamefont {M.}~\bibnamefont
			{Peev}}, \bibinfo {author} {\bibfnamefont {C.}~\bibnamefont {Pacher}},
		\bibinfo {author} {\bibfnamefont {R.}~\bibnamefont {All{\'{e}}aume}},
		\bibinfo {author} {\bibfnamefont {C.}~\bibnamefont {Barreiro}}, \bibinfo
		{author} {\bibfnamefont {J.}~\bibnamefont {Bouda}}, \bibinfo {author}
		{\bibfnamefont {W.}~\bibnamefont {Boxleitner}}, \bibinfo {author}
		{\bibfnamefont {T.}~\bibnamefont {Debuisschert}}, \bibinfo {author}
		{\bibfnamefont {E.}~\bibnamefont {Diamanti}}, \bibinfo {author}
		{\bibfnamefont {M.}~\bibnamefont {Dianati}}, \bibinfo {author} {\bibfnamefont
			{J.~F.}\ \bibnamefont {Dynes}}, \bibinfo {author} {\bibfnamefont
			{S.}~\bibnamefont {Fasel}}, \bibinfo {author} {\bibfnamefont
			{S.}~\bibnamefont {Fossier}}, \bibinfo {author} {\bibfnamefont
			{M.}~\bibnamefont {F{\"{u}}rst}}, \bibinfo {author} {\bibfnamefont {J.-D.}\
			\bibnamefont {Gautier}}, \bibinfo {author} {\bibfnamefont {O.}~\bibnamefont
			{Gay}}, \bibinfo {author} {\bibfnamefont {N.}~\bibnamefont {Gisin}}, \bibinfo
		{author} {\bibfnamefont {P.}~\bibnamefont {Grangier}}, \bibinfo {author}
		{\bibfnamefont {A.}~\bibnamefont {Happe}}, \bibinfo {author} {\bibfnamefont
			{Y.}~\bibnamefont {Hasani}}, \bibinfo {author} {\bibfnamefont
			{M.}~\bibnamefont {Hentschel}}, \bibinfo {author} {\bibfnamefont
			{H.}~\bibnamefont {H{\"{u}}bel}}, \bibinfo {author} {\bibfnamefont
			{G.}~\bibnamefont {Humer}}, \bibinfo {author} {\bibfnamefont
			{T.}~\bibnamefont {L{\"{a}}nger}}, \bibinfo {author} {\bibfnamefont
			{M.}~\bibnamefont {Legr{\'{e}}}}, \bibinfo {author} {\bibfnamefont
			{R.}~\bibnamefont {Lieger}}, \bibinfo {author} {\bibfnamefont
			{J.}~\bibnamefont {Lodewyck}}, \bibinfo {author} {\bibfnamefont
			{T.}~\bibnamefont {Lor{\"{u}}nser}}, \bibinfo {author} {\bibfnamefont
			{N.}~\bibnamefont {L{\"{u}}tkenhaus}}, \bibinfo {author} {\bibfnamefont
			{A.}~\bibnamefont {Marhold}}, \bibinfo {author} {\bibfnamefont
			{T.}~\bibnamefont {Matyus}}, \bibinfo {author} {\bibfnamefont
			{O.}~\bibnamefont {Maurhart}}, \bibinfo {author} {\bibfnamefont
			{L.}~\bibnamefont {Monat}}, \bibinfo {author} {\bibfnamefont
			{S.}~\bibnamefont {Nauerth}}, \bibinfo {author} {\bibfnamefont {J.-B.}\
			\bibnamefont {Page}}, \bibinfo {author} {\bibfnamefont {A.}~\bibnamefont
			{Poppe}}, \bibinfo {author} {\bibfnamefont {E.}~\bibnamefont {Querasser}},
		\bibinfo {author} {\bibfnamefont {G.}~\bibnamefont {Ribordy}}, \bibinfo
		{author} {\bibfnamefont {S.}~\bibnamefont {Robyr}}, \bibinfo {author}
		{\bibfnamefont {L.}~\bibnamefont {Salvail}}, \bibinfo {author} {\bibfnamefont
			{A.~W.}\ \bibnamefont {Sharpe}}, \bibinfo {author} {\bibfnamefont {A.~J.}\
			\bibnamefont {Shields}}, \bibinfo {author} {\bibfnamefont {D.}~\bibnamefont
			{Stucki}}, \bibinfo {author} {\bibfnamefont {M.}~\bibnamefont {Suda}},
		\bibinfo {author} {\bibfnamefont {C.}~\bibnamefont {Tamas}}, \bibinfo
		{author} {\bibfnamefont {T.}~\bibnamefont {Themel}}, \bibinfo {author}
		{\bibfnamefont {R.~T.}\ \bibnamefont {Thew}}, \bibinfo {author}
		{\bibfnamefont {Y.}~\bibnamefont {Thoma}}, \bibinfo {author} {\bibfnamefont
			{A.}~\bibnamefont {Treiber}}, \bibinfo {author} {\bibfnamefont
			{P.}~\bibnamefont {Trinkler}}, \bibinfo {author} {\bibfnamefont
			{R.}~\bibnamefont {Tualle-Brouri}}, \bibinfo {author} {\bibfnamefont
			{F.}~\bibnamefont {Vannel}}, \bibinfo {author} {\bibfnamefont
			{N.}~\bibnamefont {Walenta}}, \bibinfo {author} {\bibfnamefont
			{H.}~\bibnamefont {Weier}}, \bibinfo {author} {\bibfnamefont
			{H.}~\bibnamefont {Weinfurter}}, \bibinfo {author} {\bibfnamefont
			{I.}~\bibnamefont {Wimberger}}, \bibinfo {author} {\bibfnamefont {Z.~L.}\
			\bibnamefont {Yuan}}, \bibinfo {author} {\bibfnamefont {H.}~\bibnamefont
			{Zbinden}}, \ and\ \bibinfo {author} {\bibfnamefont {A.}~\bibnamefont
			{Zeilinger}},\ }\href {\doibase 10.1088/1367-2630/11/7/075001} {\bibfield
		{journal} {\bibinfo  {journal} {New Journal of Physics}\ }\textbf {\bibinfo
			{volume} {11}},\ \bibinfo {pages} {075001} (\bibinfo {year}
		{2009})}\BibitemShut {NoStop}%
	\bibitem [{\citenamefont {Xu}\ \emph {et~al.}(2009)\citenamefont {Xu},
		\citenamefont {Chen}, \citenamefont {Wang}, \citenamefont {Yin},
		\citenamefont {Zhang}, \citenamefont {Liu}, \citenamefont {Zhou},
		\citenamefont {Zhao}, \citenamefont {Li}, \citenamefont {Liu}, \citenamefont
		{Han},\ and\ \citenamefont {Guo}}]{Xu2009}%
	\BibitemOpen
	\bibfield  {author} {\bibinfo {author} {\bibfnamefont {F.}~\bibnamefont
			{Xu}}, \bibinfo {author} {\bibfnamefont {W.}~\bibnamefont {Chen}}, \bibinfo
		{author} {\bibfnamefont {S.}~\bibnamefont {Wang}}, \bibinfo {author}
		{\bibfnamefont {Z.}~\bibnamefont {Yin}}, \bibinfo {author} {\bibfnamefont
			{Y.}~\bibnamefont {Zhang}}, \bibinfo {author} {\bibfnamefont
			{Y.}~\bibnamefont {Liu}}, \bibinfo {author} {\bibfnamefont {Z.}~\bibnamefont
			{Zhou}}, \bibinfo {author} {\bibfnamefont {Y.}~\bibnamefont {Zhao}}, \bibinfo
		{author} {\bibfnamefont {H.}~\bibnamefont {Li}}, \bibinfo {author}
		{\bibfnamefont {D.}~\bibnamefont {Liu}}, \bibinfo {author} {\bibfnamefont
			{Z.}~\bibnamefont {Han}}, \ and\ \bibinfo {author} {\bibfnamefont
			{G.}~\bibnamefont {Guo}},\ }\href {\doibase 10.1007/s11434-009-0526-3}
	{\bibfield  {journal} {\bibinfo  {journal} {Chinese Science Bulletin}\
		}\textbf {\bibinfo {volume} {54}},\ \bibinfo {pages} {2991} (\bibinfo {year}
		{2009})}\BibitemShut {NoStop}%
	\bibitem [{\citenamefont {Elliott}\ \emph {et~al.}(2005)\citenamefont
		{Elliott}, \citenamefont {Colvin}, \citenamefont {Pearson}, \citenamefont
		{Pikalo}, \citenamefont {Schlafer},\ and\ \citenamefont {Yeh}}]{Elliott2005}%
	\BibitemOpen
	\bibfield  {author} {\bibinfo {author} {\bibfnamefont {C.}~\bibnamefont
			{Elliott}}, \bibinfo {author} {\bibfnamefont {A.}~\bibnamefont {Colvin}},
		\bibinfo {author} {\bibfnamefont {D.}~\bibnamefont {Pearson}}, \bibinfo
		{author} {\bibfnamefont {O.}~\bibnamefont {Pikalo}}, \bibinfo {author}
		{\bibfnamefont {J.}~\bibnamefont {Schlafer}}, \ and\ \bibinfo {author}
		{\bibfnamefont {H.}~\bibnamefont {Yeh}},\ }in\ \href {\doibase
		10.1117/12.606489} {\emph {\bibinfo {booktitle} {Proc. SPIE 5815, Quantum
				Information and Computation III,}}},\ Vol.\ \bibinfo {volume} {5815},\
	\bibinfo {editor} {edited by\ \bibinfo {editor} {\bibfnamefont {E.~J.}\
			\bibnamefont {Donkor}}, \bibinfo {editor} {\bibfnamefont {A.~R.}\
			\bibnamefont {Pirich}}, \ and\ \bibinfo {editor} {\bibfnamefont {H.~E.}\
			\bibnamefont {Brandt}}}\ (\bibinfo {year} {2005})\ pp.\ \bibinfo {pages}
	{138--149}\BibitemShut {NoStop}%
	\bibitem [{\citenamefont {Courtland}(2016)}]{courtland2016chinalink}%
	\BibitemOpen
	\bibfield  {author} {\bibinfo {author} {\bibfnamefont {R.}~\bibnamefont
			{Courtland}},\ }\href {\doibase 10.1109/MSPEC.2016.7607012} {\bibfield
		{journal} {\bibinfo  {journal} {IEEE Spectrum}\ }\textbf {\bibinfo {volume}
			{53}},\ \bibinfo {pages} {11} (\bibinfo {year} {2016})}\BibitemShut {NoStop}%
	\bibitem [{\citenamefont {Wang}(2017)}]{wang2017cas_chinalink}%
	\BibitemOpen
	\bibfield  {author} {\bibinfo {author} {\bibfnamefont {M.}~\bibnamefont
			{Wang}},\ }\href {\doibase 10.1093/nsr/nwx025} {\bibfield  {journal}
		{\bibinfo  {journal} {National Science Review}\ }\textbf {\bibinfo {volume}
			{4}},\ \bibinfo {pages} {144} (\bibinfo {year} {2017})}\BibitemShut {NoStop}%
	\bibitem [{\citenamefont {Wang}\ \emph {et~al.}(2014)\citenamefont {Wang},
		\citenamefont {Chen}, \citenamefont {Yin}, \citenamefont {Li}, \citenamefont
		{He}, \citenamefont {Li}, \citenamefont {Zhou}, \citenamefont {Song},
		\citenamefont {Li}, \citenamefont {Wang}, \citenamefont {Chen}, \citenamefont
		{Han}, \citenamefont {Huang}, \citenamefont {Guo}, \citenamefont {Hao},
		\citenamefont {Li}, \citenamefont {Zhang}, \citenamefont {Liu}, \citenamefont
		{Liang}, \citenamefont {Miao}, \citenamefont {Wu}, \citenamefont {Guo},\ and\
		\citenamefont {Han}}]{wang2014fieldtest_wuhu_hefei}%
	\BibitemOpen
	\bibfield  {author} {\bibinfo {author} {\bibfnamefont {S.}~\bibnamefont
			{Wang}}, \bibinfo {author} {\bibfnamefont {W.}~\bibnamefont {Chen}}, \bibinfo
		{author} {\bibfnamefont {Z.-Q.}\ \bibnamefont {Yin}}, \bibinfo {author}
		{\bibfnamefont {H.-W.}\ \bibnamefont {Li}}, \bibinfo {author} {\bibfnamefont
			{D.-Y.}\ \bibnamefont {He}}, \bibinfo {author} {\bibfnamefont {Y.-H.}\
			\bibnamefont {Li}}, \bibinfo {author} {\bibfnamefont {Z.}~\bibnamefont
			{Zhou}}, \bibinfo {author} {\bibfnamefont {X.-T.}\ \bibnamefont {Song}},
		\bibinfo {author} {\bibfnamefont {F.-Y.}\ \bibnamefont {Li}}, \bibinfo
		{author} {\bibfnamefont {D.}~\bibnamefont {Wang}}, \bibinfo {author}
		{\bibfnamefont {H.}~\bibnamefont {Chen}}, \bibinfo {author} {\bibfnamefont
			{Y.-G.}\ \bibnamefont {Han}}, \bibinfo {author} {\bibfnamefont {J.-Z.}\
			\bibnamefont {Huang}}, \bibinfo {author} {\bibfnamefont {J.-F.}\ \bibnamefont
			{Guo}}, \bibinfo {author} {\bibfnamefont {P.-L.}\ \bibnamefont {Hao}},
		\bibinfo {author} {\bibfnamefont {M.}~\bibnamefont {Li}}, \bibinfo {author}
		{\bibfnamefont {C.-M.}\ \bibnamefont {Zhang}}, \bibinfo {author}
		{\bibfnamefont {D.}~\bibnamefont {Liu}}, \bibinfo {author} {\bibfnamefont
			{W.-Y.}\ \bibnamefont {Liang}}, \bibinfo {author} {\bibfnamefont {C.-H.}\
			\bibnamefont {Miao}}, \bibinfo {author} {\bibfnamefont {P.}~\bibnamefont
			{Wu}}, \bibinfo {author} {\bibfnamefont {G.-C.}\ \bibnamefont {Guo}}, \ and\
		\bibinfo {author} {\bibfnamefont {Z.-F.}\ \bibnamefont {Han}},\ }\href
	{\doibase 10.1364/OE.22.021739} {\bibfield  {journal} {\bibinfo  {journal}
			{Optics Express}\ }\textbf {\bibinfo {volume} {22}},\ \bibinfo {pages}
		{21739} (\bibinfo {year} {2014})}\BibitemShut {NoStop}%
	\bibitem [{\citenamefont {Liao}\ \emph {et~al.}(2017)\citenamefont {Liao},
		\citenamefont {Cai}, \citenamefont {Liu}, \citenamefont {Zhang},
		\citenamefont {Li}, \citenamefont {Ren}, \citenamefont {Yin}, \citenamefont
		{Shen}, \citenamefont {Cao}, \citenamefont {Li}, \citenamefont {Li},
		\citenamefont {Chen}, \citenamefont {Sun}, \citenamefont {Jia}, \citenamefont
		{Wu}, \citenamefont {Jiang}, \citenamefont {Wang}, \citenamefont {Huang},
		\citenamefont {Wang}, \citenamefont {Zhou}, \citenamefont {Deng},
		\citenamefont {Xi}, \citenamefont {Ma}, \citenamefont {Hu}, \citenamefont
		{Zhang}, \citenamefont {Chen}, \citenamefont {Liu}, \citenamefont {Wang},
		\citenamefont {Zhu}, \citenamefont {Lu}, \citenamefont {Shu}, \citenamefont
		{Peng}, \citenamefont {Wang},\ and\ \citenamefont
		{Pan}}]{liao2017satelliteqkd}%
	\BibitemOpen
	\bibfield  {author} {\bibinfo {author} {\bibfnamefont {S.-K.}\ \bibnamefont
			{Liao}}, \bibinfo {author} {\bibfnamefont {W.-Q.}\ \bibnamefont {Cai}},
		\bibinfo {author} {\bibfnamefont {W.-Y.}\ \bibnamefont {Liu}}, \bibinfo
		{author} {\bibfnamefont {L.}~\bibnamefont {Zhang}}, \bibinfo {author}
		{\bibfnamefont {Y.}~\bibnamefont {Li}}, \bibinfo {author} {\bibfnamefont
			{J.-G.}\ \bibnamefont {Ren}}, \bibinfo {author} {\bibfnamefont
			{J.}~\bibnamefont {Yin}}, \bibinfo {author} {\bibfnamefont {Q.}~\bibnamefont
			{Shen}}, \bibinfo {author} {\bibfnamefont {Y.}~\bibnamefont {Cao}}, \bibinfo
		{author} {\bibfnamefont {Z.-P.}\ \bibnamefont {Li}}, \bibinfo {author}
		{\bibfnamefont {F.-Z.}\ \bibnamefont {Li}}, \bibinfo {author} {\bibfnamefont
			{X.-W.}\ \bibnamefont {Chen}}, \bibinfo {author} {\bibfnamefont {L.-H.}\
			\bibnamefont {Sun}}, \bibinfo {author} {\bibfnamefont {J.-J.}\ \bibnamefont
			{Jia}}, \bibinfo {author} {\bibfnamefont {J.-C.}\ \bibnamefont {Wu}},
		\bibinfo {author} {\bibfnamefont {X.-J.}\ \bibnamefont {Jiang}}, \bibinfo
		{author} {\bibfnamefont {J.-F.}\ \bibnamefont {Wang}}, \bibinfo {author}
		{\bibfnamefont {Y.-M.}\ \bibnamefont {Huang}}, \bibinfo {author}
		{\bibfnamefont {Q.}~\bibnamefont {Wang}}, \bibinfo {author} {\bibfnamefont
			{Y.-L.}\ \bibnamefont {Zhou}}, \bibinfo {author} {\bibfnamefont
			{L.}~\bibnamefont {Deng}}, \bibinfo {author} {\bibfnamefont {T.}~\bibnamefont
			{Xi}}, \bibinfo {author} {\bibfnamefont {L.}~\bibnamefont {Ma}}, \bibinfo
		{author} {\bibfnamefont {T.}~\bibnamefont {Hu}}, \bibinfo {author}
		{\bibfnamefont {Q.}~\bibnamefont {Zhang}}, \bibinfo {author} {\bibfnamefont
			{Y.-A.}\ \bibnamefont {Chen}}, \bibinfo {author} {\bibfnamefont {N.-L.}\
			\bibnamefont {Liu}}, \bibinfo {author} {\bibfnamefont {X.-B.}\ \bibnamefont
			{Wang}}, \bibinfo {author} {\bibfnamefont {Z.-C.}\ \bibnamefont {Zhu}},
		\bibinfo {author} {\bibfnamefont {C.-Y.}\ \bibnamefont {Lu}}, \bibinfo
		{author} {\bibfnamefont {R.}~\bibnamefont {Shu}}, \bibinfo {author}
		{\bibfnamefont {C.-Z.}\ \bibnamefont {Peng}}, \bibinfo {author}
		{\bibfnamefont {J.-Y.}\ \bibnamefont {Wang}}, \ and\ \bibinfo {author}
		{\bibfnamefont {J.-W.}\ \bibnamefont {Pan}},\ }\href {\doibase
		10.1038/nature23655} {\bibfield  {journal} {\bibinfo  {journal} {Nature}\
		}\textbf {\bibinfo {volume} {549}},\ \bibinfo {pages} {43} (\bibinfo {year}
		{2017})}\BibitemShut {NoStop}%
	\bibitem [{\citenamefont {Liao}\ \emph {et~al.}(2018)\citenamefont {Liao},
		\citenamefont {Cai}, \citenamefont {Handsteiner}, \citenamefont {Liu},
		\citenamefont {Yin}, \citenamefont {Zhang}, \citenamefont {Rauch},
		\citenamefont {Fink}, \citenamefont {Ren}, \citenamefont {Liu}, \citenamefont
		{Li}, \citenamefont {Shen}, \citenamefont {Cao}, \citenamefont {Li},
		\citenamefont {Wang}, \citenamefont {Huang}, \citenamefont {Deng},
		\citenamefont {Xi}, \citenamefont {Ma}, \citenamefont {Hu}, \citenamefont
		{Li}, \citenamefont {Liu}, \citenamefont {Koidl}, \citenamefont {Wang},
		\citenamefont {Chen}, \citenamefont {Wang}, \citenamefont {Steindorfer},
		\citenamefont {Kirchner}, \citenamefont {Lu}, \citenamefont {Shu},
		\citenamefont {Ursin}, \citenamefont {Scheidl}, \citenamefont {Peng},
		\citenamefont {Wang}, \citenamefont {Zeilinger},\ and\ \citenamefont
		{Pan}}]{Liao2018_intercontinentalqkd}%
	\BibitemOpen
	\bibfield  {author} {\bibinfo {author} {\bibfnamefont {S.-K.}\ \bibnamefont
			{Liao}}, \bibinfo {author} {\bibfnamefont {W.-Q.}\ \bibnamefont {Cai}},
		\bibinfo {author} {\bibfnamefont {J.}~\bibnamefont {Handsteiner}}, \bibinfo
		{author} {\bibfnamefont {B.}~\bibnamefont {Liu}}, \bibinfo {author}
		{\bibfnamefont {J.}~\bibnamefont {Yin}}, \bibinfo {author} {\bibfnamefont
			{L.}~\bibnamefont {Zhang}}, \bibinfo {author} {\bibfnamefont
			{D.}~\bibnamefont {Rauch}}, \bibinfo {author} {\bibfnamefont
			{M.}~\bibnamefont {Fink}}, \bibinfo {author} {\bibfnamefont {J.-G.}\
			\bibnamefont {Ren}}, \bibinfo {author} {\bibfnamefont {W.-Y.}\ \bibnamefont
			{Liu}}, \bibinfo {author} {\bibfnamefont {Y.}~\bibnamefont {Li}}, \bibinfo
		{author} {\bibfnamefont {Q.}~\bibnamefont {Shen}}, \bibinfo {author}
		{\bibfnamefont {Y.}~\bibnamefont {Cao}}, \bibinfo {author} {\bibfnamefont
			{F.-Z.}\ \bibnamefont {Li}}, \bibinfo {author} {\bibfnamefont {J.-F.}\
			\bibnamefont {Wang}}, \bibinfo {author} {\bibfnamefont {Y.-M.}\ \bibnamefont
			{Huang}}, \bibinfo {author} {\bibfnamefont {L.}~\bibnamefont {Deng}},
		\bibinfo {author} {\bibfnamefont {T.}~\bibnamefont {Xi}}, \bibinfo {author}
		{\bibfnamefont {L.}~\bibnamefont {Ma}}, \bibinfo {author} {\bibfnamefont
			{T.}~\bibnamefont {Hu}}, \bibinfo {author} {\bibfnamefont {L.}~\bibnamefont
			{Li}}, \bibinfo {author} {\bibfnamefont {N.-L.}\ \bibnamefont {Liu}},
		\bibinfo {author} {\bibfnamefont {F.}~\bibnamefont {Koidl}}, \bibinfo
		{author} {\bibfnamefont {P.}~\bibnamefont {Wang}}, \bibinfo {author}
		{\bibfnamefont {Y.-A.}\ \bibnamefont {Chen}}, \bibinfo {author}
		{\bibfnamefont {X.-B.}\ \bibnamefont {Wang}}, \bibinfo {author}
		{\bibfnamefont {M.}~\bibnamefont {Steindorfer}}, \bibinfo {author}
		{\bibfnamefont {G.}~\bibnamefont {Kirchner}}, \bibinfo {author}
		{\bibfnamefont {C.-Y.}\ \bibnamefont {Lu}}, \bibinfo {author} {\bibfnamefont
			{R.}~\bibnamefont {Shu}}, \bibinfo {author} {\bibfnamefont {R.}~\bibnamefont
			{Ursin}}, \bibinfo {author} {\bibfnamefont {T.}~\bibnamefont {Scheidl}},
		\bibinfo {author} {\bibfnamefont {C.-Z.}\ \bibnamefont {Peng}}, \bibinfo
		{author} {\bibfnamefont {J.-Y.}\ \bibnamefont {Wang}}, \bibinfo {author}
		{\bibfnamefont {A.}~\bibnamefont {Zeilinger}}, \ and\ \bibinfo {author}
		{\bibfnamefont {J.-W.}\ \bibnamefont {Pan}},\ }\href {\doibase
		10.1103/PhysRevLett.120.030501} {\bibfield  {journal} {\bibinfo  {journal}
			{Physical Review Letters}\ }\textbf {\bibinfo {volume} {120}},\ \bibinfo
		{pages} {030501} (\bibinfo {year} {2018})}\BibitemShut {NoStop}%
	\bibitem [{\citenamefont {Takenaka}\ \emph {et~al.}(2017)\citenamefont
		{Takenaka}, \citenamefont {Carrasco-Casado}, \citenamefont {Fujiwara},
		\citenamefont {Kitamura}, \citenamefont {Sasaki},\ and\ \citenamefont
		{Toyoshima}}]{takenaka2017satellite}%
	\BibitemOpen
	\bibfield  {author} {\bibinfo {author} {\bibfnamefont {H.}~\bibnamefont
			{Takenaka}}, \bibinfo {author} {\bibfnamefont {A.}~\bibnamefont
			{Carrasco-Casado}}, \bibinfo {author} {\bibfnamefont {M.}~\bibnamefont
			{Fujiwara}}, \bibinfo {author} {\bibfnamefont {M.}~\bibnamefont {Kitamura}},
		\bibinfo {author} {\bibfnamefont {M.}~\bibnamefont {Sasaki}}, \ and\ \bibinfo
		{author} {\bibfnamefont {M.}~\bibnamefont {Toyoshima}},\ }\href {\doibase
		10.1038/nphoton.2017.107} {\bibfield  {journal} {\bibinfo  {journal} {Nature
				Photonics}\ }\textbf {\bibinfo {volume} {11}},\ \bibinfo {pages} {502}
		(\bibinfo {year} {2017})}\BibitemShut {NoStop}%
	\bibitem [{\citenamefont {G{\"{u}}nthner}\ \emph {et~al.}(2016)\citenamefont
		{G{\"{u}}nthner}, \citenamefont {Khan}, \citenamefont {Elser}, \citenamefont
		{Stiller}, \citenamefont {Bayraktar}, \citenamefont {M{\"{u}}ller},
		\citenamefont {Saucke}, \citenamefont {Tr{\"{o}}ndle}, \citenamefont {Heine},
		\citenamefont {Seel}, \citenamefont {Greulich}, \citenamefont {Zech},
		\citenamefont {G{\"{u}}tlich}, \citenamefont {Philipp-May}, \citenamefont
		{Marquardt},\ and\ \citenamefont {Leuchs}}]{gunthner2017quantum}%
	\BibitemOpen
	\bibfield  {author} {\bibinfo {author} {\bibfnamefont {K.}~\bibnamefont
			{G{\"{u}}nthner}}, \bibinfo {author} {\bibfnamefont {I.}~\bibnamefont
			{Khan}}, \bibinfo {author} {\bibfnamefont {D.}~\bibnamefont {Elser}},
		\bibinfo {author} {\bibfnamefont {B.}~\bibnamefont {Stiller}}, \bibinfo
		{author} {\bibfnamefont {{\"{O}}.}~\bibnamefont {Bayraktar}}, \bibinfo
		{author} {\bibfnamefont {C.~R.}\ \bibnamefont {M{\"{u}}ller}}, \bibinfo
		{author} {\bibfnamefont {K.}~\bibnamefont {Saucke}}, \bibinfo {author}
		{\bibfnamefont {D.}~\bibnamefont {Tr{\"{o}}ndle}}, \bibinfo {author}
		{\bibfnamefont {F.}~\bibnamefont {Heine}}, \bibinfo {author} {\bibfnamefont
			{S.}~\bibnamefont {Seel}}, \bibinfo {author} {\bibfnamefont {P.}~\bibnamefont
			{Greulich}}, \bibinfo {author} {\bibfnamefont {H.}~\bibnamefont {Zech}},
		\bibinfo {author} {\bibfnamefont {B.}~\bibnamefont {G{\"{u}}tlich}}, \bibinfo
		{author} {\bibfnamefont {S.}~\bibnamefont {Philipp-May}}, \bibinfo {author}
		{\bibfnamefont {C.}~\bibnamefont {Marquardt}}, \ and\ \bibinfo {author}
		{\bibfnamefont {G.}~\bibnamefont {Leuchs}},\ }\href {\doibase
		10.1364/OPTICA.4.000611} {\bibfield  {journal} {\bibinfo  {journal} {Optica}\
		}\textbf {\bibinfo {volume} {4}},\ \bibinfo {pages} {611} (\bibinfo {year}
		{2016})}\BibitemShut {NoStop}%
	\bibitem [{\citenamefont {Masanes}\ \emph {et~al.}(2011)\citenamefont
		{Masanes}, \citenamefont {Pironio},\ and\ \citenamefont
		{Ac{\'{i}}n}}]{masanes2011_device_indep}%
	\BibitemOpen
	\bibfield  {author} {\bibinfo {author} {\bibfnamefont {L.}~\bibnamefont
			{Masanes}}, \bibinfo {author} {\bibfnamefont {S.}~\bibnamefont {Pironio}}, \
		and\ \bibinfo {author} {\bibfnamefont {A.}~\bibnamefont {Ac{\'{i}}n}},\
	}\href {\doibase 10.1038/ncomms1244} {\bibfield  {journal} {\bibinfo
			{journal} {Nature Communications}\ }\textbf {\bibinfo {volume} {2}},\
		\bibinfo {pages} {238} (\bibinfo {year} {2011})}\BibitemShut {NoStop}%
	\bibitem [{\citenamefont {Branciard}\ \emph {et~al.}(2012)\citenamefont
		{Branciard}, \citenamefont {Cavalcanti}, \citenamefont {Walborn},
		\citenamefont {Scarani},\ and\ \citenamefont
		{Wiseman}}]{branciard2012_semi_device_indep}%
	\BibitemOpen
	\bibfield  {author} {\bibinfo {author} {\bibfnamefont {C.}~\bibnamefont
			{Branciard}}, \bibinfo {author} {\bibfnamefont {E.~G.}\ \bibnamefont
			{Cavalcanti}}, \bibinfo {author} {\bibfnamefont {S.~P.}\ \bibnamefont
			{Walborn}}, \bibinfo {author} {\bibfnamefont {V.}~\bibnamefont {Scarani}}, \
		and\ \bibinfo {author} {\bibfnamefont {H.~M.}\ \bibnamefont {Wiseman}},\
	}\href {\doibase 10.1103/PhysRevA.85.010301} {\bibfield  {journal} {\bibinfo
			{journal} {Physical Review A}\ }\textbf {\bibinfo {volume} {85}},\ \bibinfo
		{pages} {010301} (\bibinfo {year} {2012})}\BibitemShut {NoStop}%
	\bibitem [{\citenamefont {Marcikic}\ \emph {et~al.}(2002)\citenamefont
		{Marcikic}, \citenamefont {de~Riedmatten}, \citenamefont {Tittel},
		\citenamefont {Scarani}, \citenamefont {Zbinden},\ and\ \citenamefont
		{Gisin}}]{Marcikic2002}%
	\BibitemOpen
	\bibfield  {author} {\bibinfo {author} {\bibfnamefont {I.}~\bibnamefont
			{Marcikic}}, \bibinfo {author} {\bibfnamefont {H.}~\bibnamefont
			{de~Riedmatten}}, \bibinfo {author} {\bibfnamefont {W.}~\bibnamefont
			{Tittel}}, \bibinfo {author} {\bibfnamefont {V.}~\bibnamefont {Scarani}},
		\bibinfo {author} {\bibfnamefont {H.}~\bibnamefont {Zbinden}}, \ and\
		\bibinfo {author} {\bibfnamefont {N.}~\bibnamefont {Gisin}},\ }\href
	{\doibase 10.1103/PhysRevA.66.062308} {\bibfield  {journal} {\bibinfo
			{journal} {Physical Review A}\ }\textbf {\bibinfo {volume} {66}},\ \bibinfo
		{pages} {062308} (\bibinfo {year} {2002})}\BibitemShut {NoStop}%
	\bibitem [{\citenamefont {Gisin}\ \emph {et~al.}(2002)\citenamefont {Gisin},
		\citenamefont {Ribordy}, \citenamefont {Tittel},\ and\ \citenamefont
		{Zbinden}}]{gisin2002}%
	\BibitemOpen
	\bibfield  {author} {\bibinfo {author} {\bibfnamefont {N.}~\bibnamefont
			{Gisin}}, \bibinfo {author} {\bibfnamefont {G.}~\bibnamefont {Ribordy}},
		\bibinfo {author} {\bibfnamefont {W.}~\bibnamefont {Tittel}}, \ and\ \bibinfo
		{author} {\bibfnamefont {H.}~\bibnamefont {Zbinden}},\ }\href {\doibase
		10.1103/RevModPhys.74.145} {\bibfield  {journal} {\bibinfo  {journal}
			{Reviews of Modern Physics}\ }\textbf {\bibinfo {volume} {74}},\ \bibinfo
		{pages} {145} (\bibinfo {year} {2002})}\BibitemShut {NoStop}%
	\bibitem [{\citenamefont {Brodsky}\ \emph {et~al.}(2011)\citenamefont
		{Brodsky}, \citenamefont {George}, \citenamefont {Antonelli},\ and\
		\citenamefont {Shtaif}}]{brodsky2011loss}%
	\BibitemOpen
	\bibfield  {author} {\bibinfo {author} {\bibfnamefont {M.}~\bibnamefont
			{Brodsky}}, \bibinfo {author} {\bibfnamefont {E.~C.}\ \bibnamefont {George}},
		\bibinfo {author} {\bibfnamefont {C.}~\bibnamefont {Antonelli}}, \ and\
		\bibinfo {author} {\bibfnamefont {M.}~\bibnamefont {Shtaif}},\ }\href
	{\doibase 10.1364/OL.36.000043} {\bibfield  {journal} {\bibinfo  {journal}
			{Optics Letters}\ }\textbf {\bibinfo {volume} {36}},\ \bibinfo {pages} {43}
		(\bibinfo {year} {2011})}\BibitemShut {NoStop}%
	\bibitem [{\citenamefont {Antonelli}\ \emph {et~al.}(2011)\citenamefont
		{Antonelli}, \citenamefont {Shtaif},\ and\ \citenamefont
		{Brodsky}}]{antonelli2011sudden}%
	\BibitemOpen
	\bibfield  {author} {\bibinfo {author} {\bibfnamefont {C.}~\bibnamefont
			{Antonelli}}, \bibinfo {author} {\bibfnamefont {M.}~\bibnamefont {Shtaif}}, \
		and\ \bibinfo {author} {\bibfnamefont {M.}~\bibnamefont {Brodsky}},\ }\href
	{http://dx.doi.org/10.1103/PhysRevLett.106.080404} {\bibfield  {journal}
		{\bibinfo  {journal} {Phys. Rev. Lett}\ }\textbf {\bibinfo {volume} {106}},\
		\bibinfo {pages} {80404} (\bibinfo {year} {2011})}\BibitemShut {NoStop}%
	\bibitem [{\citenamefont {Boller}(2016{\natexlab{a}})}]{NASAMap1}%
	\BibitemOpen
	\bibfield  {author} {\bibinfo {author} {\bibfnamefont {R.}~\bibnamefont
			{Boller}},\ }\href
	{https://worldview.earthdata.nasa.gov/?p=antarctic{\&}l=MODIS{\_}Aqua{\_}CorrectedReflectance{\_}TrueColor,MODIS{\_}Terra{\_}CorrectedReflectance{\_}TrueColor,Coastlines{\&}t=2012-07-24{\&}v=97132.18131775898,-1410839.8154507237,446828.181317759,-1246231.8154507237}
	{\enquote {\bibinfo {title} {{NASA Worldview}},}\ } (\bibinfo {year}
	{2016}{\natexlab{a}})\BibitemShut {NoStop}%
	\bibitem [{\citenamefont {Boller}(2016{\natexlab{b}})}]{NASAMap2}%
	\BibitemOpen
	\bibfield  {author} {\bibinfo {author} {\bibfnamefont {R.}~\bibnamefont
			{Boller}},\ }\href
	{https://worldview.earthdata.nasa.gov/?p=antarctic{\&}l=MODIS{\_}Aqua{\_}CorrectedReflectance{\_}TrueColor,MODIS{\_}Terra{\_}CorrectedReflectance{\_}TrueColor,Coastlines{\&}t=2012-07-24{\&}v=97132.18131775898,-1410839.8154507237,446828.181317759,-1246231.8154507237}
	{\enquote {\bibinfo {title} {{NASA Worldview}},}\ } (\bibinfo {year}
	{2016}{\natexlab{b}})\BibitemShut {NoStop}%
	\bibitem [{\citenamefont {Yin}\ \emph {et~al.}(2012)\citenamefont {Yin},
		\citenamefont {Ren}, \citenamefont {Lu}, \citenamefont {Cao}, \citenamefont
		{Yong}, \citenamefont {Wu}, \citenamefont {Liu}, \citenamefont {Liao},
		\citenamefont {Zhou}, \citenamefont {Jiang}, \citenamefont {Cai},
		\citenamefont {Xu}, \citenamefont {Pan}, \citenamefont {Jia}, \citenamefont
		{Huang}, \citenamefont {Yin}, \citenamefont {Wang}, \citenamefont {Chen},
		\citenamefont {Peng},\ and\ \citenamefont {Pan}}]{yin2012_lake}%
	\BibitemOpen
	\bibfield  {author} {\bibinfo {author} {\bibfnamefont {J.}~\bibnamefont
			{Yin}}, \bibinfo {author} {\bibfnamefont {J.-G.}\ \bibnamefont {Ren}},
		\bibinfo {author} {\bibfnamefont {H.}~\bibnamefont {Lu}}, \bibinfo {author}
		{\bibfnamefont {Y.}~\bibnamefont {Cao}}, \bibinfo {author} {\bibfnamefont
			{H.-L.}\ \bibnamefont {Yong}}, \bibinfo {author} {\bibfnamefont {Y.-P.}\
			\bibnamefont {Wu}}, \bibinfo {author} {\bibfnamefont {C.}~\bibnamefont
			{Liu}}, \bibinfo {author} {\bibfnamefont {S.-K.}\ \bibnamefont {Liao}},
		\bibinfo {author} {\bibfnamefont {F.}~\bibnamefont {Zhou}}, \bibinfo {author}
		{\bibfnamefont {Y.}~\bibnamefont {Jiang}}, \bibinfo {author} {\bibfnamefont
			{X.-D.}\ \bibnamefont {Cai}}, \bibinfo {author} {\bibfnamefont
			{P.}~\bibnamefont {Xu}}, \bibinfo {author} {\bibfnamefont {G.-S.}\
			\bibnamefont {Pan}}, \bibinfo {author} {\bibfnamefont {J.-J.}\ \bibnamefont
			{Jia}}, \bibinfo {author} {\bibfnamefont {Y.-M.}\ \bibnamefont {Huang}},
		\bibinfo {author} {\bibfnamefont {H.}~\bibnamefont {Yin}}, \bibinfo {author}
		{\bibfnamefont {J.-Y.}\ \bibnamefont {Wang}}, \bibinfo {author}
		{\bibfnamefont {Y.-A.}\ \bibnamefont {Chen}}, \bibinfo {author}
		{\bibfnamefont {C.-Z.}\ \bibnamefont {Peng}}, \ and\ \bibinfo {author}
		{\bibfnamefont {J.-W.}\ \bibnamefont {Pan}},\ }\href {\doibase
		10.1038/nature11332} {\bibfield  {journal} {\bibinfo  {journal} {Nature}\
		}\textbf {\bibinfo {volume} {488}},\ \bibinfo {pages} {185} (\bibinfo {year}
		{2012})}\BibitemShut {NoStop}%
	\bibitem [{\citenamefont {Marcikic}\ \emph {et~al.}(2006)\citenamefont
		{Marcikic}, \citenamefont {Lamas-Linares},\ and\ \citenamefont
		{Kurtsiefer}}]{marcikic2006freespacekurtsiefer}%
	\BibitemOpen
	\bibfield  {author} {\bibinfo {author} {\bibfnamefont {I.}~\bibnamefont
			{Marcikic}}, \bibinfo {author} {\bibfnamefont {A.}~\bibnamefont
			{Lamas-Linares}}, \ and\ \bibinfo {author} {\bibfnamefont {C.}~\bibnamefont
			{Kurtsiefer}},\ }\href {\doibase 10.1063/1.2348775} {\bibfield  {journal}
		{\bibinfo  {journal} {Applied Physics Letters}\ }\textbf {\bibinfo {volume}
			{89}},\ \bibinfo {pages} {101122} (\bibinfo {year} {2006})}\BibitemShut
	{NoStop}%
	\bibitem [{\citenamefont {Yin}\ \emph {et~al.}(2017)\citenamefont {Yin},
		\citenamefont {Cao}, \citenamefont {Li}, \citenamefont {Liao}, \citenamefont
		{Zhang}, \citenamefont {Ren}, \citenamefont {Cai}, \citenamefont {Liu},
		\citenamefont {Li}, \citenamefont {Dai}, \citenamefont {Li}, \citenamefont
		{Lu}, \citenamefont {Gong}, \citenamefont {Xu}, \citenamefont {Li},
		\citenamefont {Li}, \citenamefont {Yin}, \citenamefont {Jiang}, \citenamefont
		{Li}, \citenamefont {Jia}, \citenamefont {Ren}, \citenamefont {He},
		\citenamefont {Zhou}, \citenamefont {Zhang}, \citenamefont {Wang},
		\citenamefont {Chang}, \citenamefont {Zhu}, \citenamefont {Liu},
		\citenamefont {Chen}, \citenamefont {Lu}, \citenamefont {Shu}, \citenamefont
		{Peng}, \citenamefont {Wang},\ and\ \citenamefont {Pan}}]{Yin2017a}%
	\BibitemOpen
	\bibfield  {author} {\bibinfo {author} {\bibfnamefont {J.}~\bibnamefont
			{Yin}}, \bibinfo {author} {\bibfnamefont {Y.}~\bibnamefont {Cao}}, \bibinfo
		{author} {\bibfnamefont {Y.-H.}\ \bibnamefont {Li}}, \bibinfo {author}
		{\bibfnamefont {S.-K.}\ \bibnamefont {Liao}}, \bibinfo {author}
		{\bibfnamefont {L.}~\bibnamefont {Zhang}}, \bibinfo {author} {\bibfnamefont
			{J.-G.}\ \bibnamefont {Ren}}, \bibinfo {author} {\bibfnamefont {W.-Q.}\
			\bibnamefont {Cai}}, \bibinfo {author} {\bibfnamefont {W.-Y.}\ \bibnamefont
			{Liu}}, \bibinfo {author} {\bibfnamefont {B.}~\bibnamefont {Li}}, \bibinfo
		{author} {\bibfnamefont {H.}~\bibnamefont {Dai}}, \bibinfo {author}
		{\bibfnamefont {G.-B.}\ \bibnamefont {Li}}, \bibinfo {author} {\bibfnamefont
			{Q.-M.}\ \bibnamefont {Lu}}, \bibinfo {author} {\bibfnamefont {Y.-H.}\
			\bibnamefont {Gong}}, \bibinfo {author} {\bibfnamefont {Y.}~\bibnamefont
			{Xu}}, \bibinfo {author} {\bibfnamefont {S.-L.}\ \bibnamefont {Li}}, \bibinfo
		{author} {\bibfnamefont {F.-Z.}\ \bibnamefont {Li}}, \bibinfo {author}
		{\bibfnamefont {Y.-Y.}\ \bibnamefont {Yin}}, \bibinfo {author} {\bibfnamefont
			{Z.-Q.}\ \bibnamefont {Jiang}}, \bibinfo {author} {\bibfnamefont
			{M.}~\bibnamefont {Li}}, \bibinfo {author} {\bibfnamefont {J.-J.}\
			\bibnamefont {Jia}}, \bibinfo {author} {\bibfnamefont {G.}~\bibnamefont
			{Ren}}, \bibinfo {author} {\bibfnamefont {D.}~\bibnamefont {He}}, \bibinfo
		{author} {\bibfnamefont {Y.-L.}\ \bibnamefont {Zhou}}, \bibinfo {author}
		{\bibfnamefont {X.-X.}\ \bibnamefont {Zhang}}, \bibinfo {author}
		{\bibfnamefont {N.}~\bibnamefont {Wang}}, \bibinfo {author} {\bibfnamefont
			{X.}~\bibnamefont {Chang}}, \bibinfo {author} {\bibfnamefont {Z.-C.}\
			\bibnamefont {Zhu}}, \bibinfo {author} {\bibfnamefont {N.-L.}\ \bibnamefont
			{Liu}}, \bibinfo {author} {\bibfnamefont {Y.-A.}\ \bibnamefont {Chen}},
		\bibinfo {author} {\bibfnamefont {C.-Y.}\ \bibnamefont {Lu}}, \bibinfo
		{author} {\bibfnamefont {R.}~\bibnamefont {Shu}}, \bibinfo {author}
		{\bibfnamefont {C.-Z.}\ \bibnamefont {Peng}}, \bibinfo {author}
		{\bibfnamefont {J.-Y.}\ \bibnamefont {Wang}}, \ and\ \bibinfo {author}
		{\bibfnamefont {J.-W.}\ \bibnamefont {Pan}},\ }\href {\doibase
		10.1126/science.aan3211} {\bibfield  {journal} {\bibinfo  {journal}
			{Science}\ }\textbf {\bibinfo {volume} {356}},\ \bibinfo {pages} {1140}
		(\bibinfo {year} {2017})}\BibitemShut {NoStop}%
	\bibitem [{\citenamefont {Ren}\ \emph {et~al.}(2017)\citenamefont {Ren},
		\citenamefont {Xu}, \citenamefont {Yong}, \citenamefont {Zhang},
		\citenamefont {Liao}, \citenamefont {Yin}, \citenamefont {Liu}, \citenamefont
		{Cai}, \citenamefont {Yang}, \citenamefont {Li}, \citenamefont {Yang},
		\citenamefont {Han}, \citenamefont {Yao}, \citenamefont {Li}, \citenamefont
		{Wu}, \citenamefont {Wan}, \citenamefont {Liu}, \citenamefont {Liu},
		\citenamefont {Kuang}, \citenamefont {He}, \citenamefont {Shang},
		\citenamefont {Guo}, \citenamefont {Zheng}, \citenamefont {Tian},
		\citenamefont {Zhu}, \citenamefont {Liu}, \citenamefont {Lu}, \citenamefont
		{Shu}, \citenamefont {Chen}, \citenamefont {Peng}, \citenamefont {Wang},\
		and\ \citenamefont {Pan}}]{ren2017_ground_to_satellite_teleportation}%
	\BibitemOpen
	\bibfield  {author} {\bibinfo {author} {\bibfnamefont {J.-G.}\ \bibnamefont
			{Ren}}, \bibinfo {author} {\bibfnamefont {P.}~\bibnamefont {Xu}}, \bibinfo
		{author} {\bibfnamefont {H.-L.}\ \bibnamefont {Yong}}, \bibinfo {author}
		{\bibfnamefont {L.}~\bibnamefont {Zhang}}, \bibinfo {author} {\bibfnamefont
			{S.-K.}\ \bibnamefont {Liao}}, \bibinfo {author} {\bibfnamefont
			{J.}~\bibnamefont {Yin}}, \bibinfo {author} {\bibfnamefont {W.-Y.}\
			\bibnamefont {Liu}}, \bibinfo {author} {\bibfnamefont {W.-Q.}\ \bibnamefont
			{Cai}}, \bibinfo {author} {\bibfnamefont {M.}~\bibnamefont {Yang}}, \bibinfo
		{author} {\bibfnamefont {L.}~\bibnamefont {Li}}, \bibinfo {author}
		{\bibfnamefont {K.-X.}\ \bibnamefont {Yang}}, \bibinfo {author}
		{\bibfnamefont {X.}~\bibnamefont {Han}}, \bibinfo {author} {\bibfnamefont
			{Y.-Q.}\ \bibnamefont {Yao}}, \bibinfo {author} {\bibfnamefont
			{J.}~\bibnamefont {Li}}, \bibinfo {author} {\bibfnamefont {H.-Y.}\
			\bibnamefont {Wu}}, \bibinfo {author} {\bibfnamefont {S.}~\bibnamefont
			{Wan}}, \bibinfo {author} {\bibfnamefont {L.}~\bibnamefont {Liu}}, \bibinfo
		{author} {\bibfnamefont {D.-Q.}\ \bibnamefont {Liu}}, \bibinfo {author}
		{\bibfnamefont {Y.-W.}\ \bibnamefont {Kuang}}, \bibinfo {author}
		{\bibfnamefont {Z.-P.}\ \bibnamefont {He}}, \bibinfo {author} {\bibfnamefont
			{P.}~\bibnamefont {Shang}}, \bibinfo {author} {\bibfnamefont
			{C.}~\bibnamefont {Guo}}, \bibinfo {author} {\bibfnamefont {R.-H.}\
			\bibnamefont {Zheng}}, \bibinfo {author} {\bibfnamefont {K.}~\bibnamefont
			{Tian}}, \bibinfo {author} {\bibfnamefont {Z.-C.}\ \bibnamefont {Zhu}},
		\bibinfo {author} {\bibfnamefont {N.-L.}\ \bibnamefont {Liu}}, \bibinfo
		{author} {\bibfnamefont {C.-Y.}\ \bibnamefont {Lu}}, \bibinfo {author}
		{\bibfnamefont {R.}~\bibnamefont {Shu}}, \bibinfo {author} {\bibfnamefont
			{Y.-A.}\ \bibnamefont {Chen}}, \bibinfo {author} {\bibfnamefont {C.-Z.}\
			\bibnamefont {Peng}}, \bibinfo {author} {\bibfnamefont {J.-Y.}\ \bibnamefont
			{Wang}}, \ and\ \bibinfo {author} {\bibfnamefont {J.-W.}\ \bibnamefont
			{Pan}},\ }\href {\doibase 10.1038/nature23675} {\bibfield  {journal}
		{\bibinfo  {journal} {Nature}\ }\textbf {\bibinfo {volume} {549}},\ \bibinfo
		{pages} {70} (\bibinfo {year} {2017})}\BibitemShut {NoStop}%
	\bibitem [{\citenamefont {H{\"{u}}bel}\ \emph {et~al.}(2008)\citenamefont
		{H{\"{u}}bel}, \citenamefont {Vanner}, \citenamefont {Lederer}, \citenamefont
		{Blauensteiner}, \citenamefont {Lor{\"{u}}nser}, \citenamefont {Poppe},\ and\
		\citenamefont {Zeilinger}}]{hubel2007high}%
	\BibitemOpen
	\bibfield  {author} {\bibinfo {author} {\bibfnamefont {H.}~\bibnamefont
			{H{\"{u}}bel}}, \bibinfo {author} {\bibfnamefont {M.~R.}\ \bibnamefont
			{Vanner}}, \bibinfo {author} {\bibfnamefont {T.}~\bibnamefont {Lederer}},
		\bibinfo {author} {\bibfnamefont {B.}~\bibnamefont {Blauensteiner}}, \bibinfo
		{author} {\bibfnamefont {T.}~\bibnamefont {Lor{\"{u}}nser}}, \bibinfo
		{author} {\bibfnamefont {A.}~\bibnamefont {Poppe}}, \ and\ \bibinfo {author}
		{\bibfnamefont {A.}~\bibnamefont {Zeilinger}},\ }\href {\doibase
		10.1364/OE.15.007853} {\bibfield  {journal} {\bibinfo  {journal} {Optics
				Express}\ }\textbf {\bibinfo {volume} {15}},\ \bibinfo {pages} {7853}
		(\bibinfo {year} {2008})}\BibitemShut {NoStop}%
	\bibitem [{\citenamefont {Inagaki}\ \emph {et~al.}(2013)\citenamefont
		{Inagaki}, \citenamefont {Matsuda}, \citenamefont {Tadanaga}, \citenamefont
		{Asobe},\ and\ \citenamefont {Takesue}}]{Inagaki2013}%
	\BibitemOpen
	\bibfield  {author} {\bibinfo {author} {\bibfnamefont {T.}~\bibnamefont
			{Inagaki}}, \bibinfo {author} {\bibfnamefont {N.}~\bibnamefont {Matsuda}},
		\bibinfo {author} {\bibfnamefont {O.}~\bibnamefont {Tadanaga}}, \bibinfo
		{author} {\bibfnamefont {M.}~\bibnamefont {Asobe}}, \ and\ \bibinfo {author}
		{\bibfnamefont {H.}~\bibnamefont {Takesue}},\ }\href {\doibase
		10.1364/OE.21.023241} {\bibfield  {journal} {\bibinfo  {journal} {Optics
				Express}\ }\textbf {\bibinfo {volume} {21}},\ \bibinfo {pages} {23241}
		(\bibinfo {year} {2013})}\BibitemShut {NoStop}%
	\bibitem [{\citenamefont {Aktas}\ \emph {et~al.}(2016)\citenamefont {Aktas},
		\citenamefont {Fedrici}, \citenamefont {Kaiser}, \citenamefont {Lunghi},
		\citenamefont {Labont{\'{e}}},\ and\ \citenamefont {Tanzilli}}]{Aktas2016}%
	\BibitemOpen
	\bibfield  {author} {\bibinfo {author} {\bibfnamefont {D.}~\bibnamefont
			{Aktas}}, \bibinfo {author} {\bibfnamefont {B.}~\bibnamefont {Fedrici}},
		\bibinfo {author} {\bibfnamefont {F.}~\bibnamefont {Kaiser}}, \bibinfo
		{author} {\bibfnamefont {T.}~\bibnamefont {Lunghi}}, \bibinfo {author}
		{\bibfnamefont {L.}~\bibnamefont {Labont{\'{e}}}}, \ and\ \bibinfo {author}
		{\bibfnamefont {S.}~\bibnamefont {Tanzilli}},\ }\href {\doibase
		10.1002/lpor.201500258} {\bibfield  {journal} {\bibinfo  {journal} {Laser
				{\&} Photonics Reviews}\ }\textbf {\bibinfo {volume} {10}},\ \bibinfo {pages}
		{451} (\bibinfo {year} {2016})}\BibitemShut {NoStop}%
	\bibitem [{\citenamefont {Honjo}\ \emph {et~al.}(2007)\citenamefont {Honjo},
		\citenamefont {Takesue}, \citenamefont {Kamada}, \citenamefont {Nishida},
		\citenamefont {Tadanaga}, \citenamefont {Asobe},\ and\ \citenamefont
		{Inoue}}]{honjo2007_100km_timebin}%
	\BibitemOpen
	\bibfield  {author} {\bibinfo {author} {\bibfnamefont {T.}~\bibnamefont
			{Honjo}}, \bibinfo {author} {\bibfnamefont {H.}~\bibnamefont {Takesue}},
		\bibinfo {author} {\bibfnamefont {H.}~\bibnamefont {Kamada}}, \bibinfo
		{author} {\bibfnamefont {Y.}~\bibnamefont {Nishida}}, \bibinfo {author}
		{\bibfnamefont {O.}~\bibnamefont {Tadanaga}}, \bibinfo {author}
		{\bibfnamefont {M.}~\bibnamefont {Asobe}}, \ and\ \bibinfo {author}
		{\bibfnamefont {K.}~\bibnamefont {Inoue}},\ }\href {\doibase
		10.1364/OE.15.013957} {\bibfield  {journal} {\bibinfo  {journal} {Optics
				Express}\ }\textbf {\bibinfo {volume} {15}},\ \bibinfo {pages} {13957}
		(\bibinfo {year} {2007})}\BibitemShut {NoStop}%
	\bibitem [{\citenamefont {Salart}\ \emph {et~al.}(2008)\citenamefont {Salart},
		\citenamefont {Baas}, \citenamefont {Branciard}, \citenamefont {Gisin},\ and\
		\citenamefont {Zbinden}}]{salart2008testing}%
	\BibitemOpen
	\bibfield  {author} {\bibinfo {author} {\bibfnamefont {D.}~\bibnamefont
			{Salart}}, \bibinfo {author} {\bibfnamefont {A.}~\bibnamefont {Baas}},
		\bibinfo {author} {\bibfnamefont {C.}~\bibnamefont {Branciard}}, \bibinfo
		{author} {\bibfnamefont {N.}~\bibnamefont {Gisin}}, \ and\ \bibinfo {author}
		{\bibfnamefont {H.}~\bibnamefont {Zbinden}},\ }\href {\doibase
		10.1038/nature07121} {\bibfield  {journal} {\bibinfo  {journal} {Nature}\
		}\textbf {\bibinfo {volume} {454}},\ \bibinfo {pages} {861} (\bibinfo {year}
		{2008})}\BibitemShut {NoStop}%
	\bibitem [{\citenamefont {Fedrizzi}\ \emph {et~al.}(2005)\citenamefont
		{Fedrizzi}, \citenamefont {Poppe}, \citenamefont {Ursin}, \citenamefont
		{Loriinser}, \citenamefont {Peev}, \citenamefont {Linger},\ and\
		\citenamefont {Zeilinger}}]{poppe2004practical}%
	\BibitemOpen
	\bibfield  {author} {\bibinfo {author} {\bibfnamefont {A.}~\bibnamefont
			{Fedrizzi}}, \bibinfo {author} {\bibfnamefont {A.}~\bibnamefont {Poppe}},
		\bibinfo {author} {\bibfnamefont {R.}~\bibnamefont {Ursin}}, \bibinfo
		{author} {\bibfnamefont {T.}~\bibnamefont {Loriinser}}, \bibinfo {author}
		{\bibfnamefont {M.}~\bibnamefont {Peev}}, \bibinfo {author} {\bibfnamefont
			{T.}~\bibnamefont {Linger}}, \ and\ \bibinfo {author} {\bibfnamefont
			{A.}~\bibnamefont {Zeilinger}},\ }\href {\doibase 10.1109/EQEC.2005.1567469}
	{\bibfield  {journal} {\bibinfo  {journal} {EQEC '05. European Quantum
				Electronics Conference, 2005.}\ }\textbf {\bibinfo {volume} {2005}},\
		\bibinfo {pages} {303} (\bibinfo {year} {2005})}\BibitemShut {NoStop}%
	\bibitem [{\citenamefont {Treiber}\ \emph {et~al.}(2009)\citenamefont
		{Treiber}, \citenamefont {Poppe}, \citenamefont {Hentschel}, \citenamefont
		{Ferrini}, \citenamefont {Lor{\"{u}}nser}, \citenamefont {Querasser},
		\citenamefont {Matyus}, \citenamefont {H{\"{u}}bel},\ and\ \citenamefont
		{Zeilinger}}]{treiber2009fully}%
	\BibitemOpen
	\bibfield  {author} {\bibinfo {author} {\bibfnamefont {A.}~\bibnamefont
			{Treiber}}, \bibinfo {author} {\bibfnamefont {A.}~\bibnamefont {Poppe}},
		\bibinfo {author} {\bibfnamefont {M.}~\bibnamefont {Hentschel}}, \bibinfo
		{author} {\bibfnamefont {D.}~\bibnamefont {Ferrini}}, \bibinfo {author}
		{\bibfnamefont {T.}~\bibnamefont {Lor{\"{u}}nser}}, \bibinfo {author}
		{\bibfnamefont {E.}~\bibnamefont {Querasser}}, \bibinfo {author}
		{\bibfnamefont {T.}~\bibnamefont {Matyus}}, \bibinfo {author} {\bibfnamefont
			{H.}~\bibnamefont {H{\"{u}}bel}}, \ and\ \bibinfo {author} {\bibfnamefont
			{A.}~\bibnamefont {Zeilinger}},\ }\href {\doibase
		10.1088/1367-2630/11/4/045013} {\bibfield  {journal} {\bibinfo  {journal}
			{New Journal of Physics}\ }\textbf {\bibinfo {volume} {11}},\ \bibinfo
		{pages} {045013} (\bibinfo {year} {2009})}\BibitemShut {NoStop}%
	\bibitem [{\citenamefont {Valivarthi}\ \emph {et~al.}(2016)\citenamefont
		{Valivarthi}, \citenamefont {Puigibert}, \citenamefont {Zhou}, \citenamefont
		{Aguilar}, \citenamefont {Verma}, \citenamefont {Marsili}, \citenamefont
		{Shaw}, \citenamefont {Nam}, \citenamefont {Oblak},\ and\ \citenamefont
		{Tittel}}]{valivarthi2016telep}%
	\BibitemOpen
	\bibfield  {author} {\bibinfo {author} {\bibfnamefont {R.}~\bibnamefont
			{Valivarthi}}, \bibinfo {author} {\bibfnamefont {M.~l.~G.}\ \bibnamefont
			{Puigibert}}, \bibinfo {author} {\bibfnamefont {Q.}~\bibnamefont {Zhou}},
		\bibinfo {author} {\bibfnamefont {G.~H.}\ \bibnamefont {Aguilar}}, \bibinfo
		{author} {\bibfnamefont {V.~B.}\ \bibnamefont {Verma}}, \bibinfo {author}
		{\bibfnamefont {F.}~\bibnamefont {Marsili}}, \bibinfo {author} {\bibfnamefont
			{M.~D.}\ \bibnamefont {Shaw}}, \bibinfo {author} {\bibfnamefont {S.~W.}\
			\bibnamefont {Nam}}, \bibinfo {author} {\bibfnamefont {D.}~\bibnamefont
			{Oblak}}, \ and\ \bibinfo {author} {\bibfnamefont {W.}~\bibnamefont
			{Tittel}},\ }\href {\doibase 10.1038/nphoton.2016.180} {\bibfield  {journal}
		{\bibinfo  {journal} {Nature Photonics}\ }\textbf {\bibinfo {volume} {10}},\
		\bibinfo {pages} {676} (\bibinfo {year} {2016})}\BibitemShut {NoStop}%
	\bibitem [{\citenamefont {Sun}\ \emph {et~al.}(2016)\citenamefont {Sun},
		\citenamefont {Mao}, \citenamefont {Chen}, \citenamefont {Zhang},
		\citenamefont {Jiang}, \citenamefont {Zhang}, \citenamefont {Zhang},
		\citenamefont {Miki}, \citenamefont {Yamashita}, \citenamefont {Terai},
		\citenamefont {Jiang}, \citenamefont {Chen}, \citenamefont {You},
		\citenamefont {Chen}, \citenamefont {Wang}, \citenamefont {Fan},
		\citenamefont {Zhang},\ and\ \citenamefont
		{Pan}}]{sun2016_chinese-teleportation-deployed-fiber}%
	\BibitemOpen
	\bibfield  {author} {\bibinfo {author} {\bibfnamefont {Q.}~\bibnamefont
			{Sun}}, \bibinfo {author} {\bibfnamefont {Y.}~\bibnamefont {Mao}}, \bibinfo
		{author} {\bibfnamefont {S.-J.}\ \bibnamefont {Chen}}, \bibinfo {author}
		{\bibfnamefont {W.}~\bibnamefont {Zhang}}, \bibinfo {author} {\bibfnamefont
			{Y.-F.}\ \bibnamefont {Jiang}}, \bibinfo {author} {\bibfnamefont {Y.-B.}\
			\bibnamefont {Zhang}}, \bibinfo {author} {\bibfnamefont {W.-J.}\ \bibnamefont
			{Zhang}}, \bibinfo {author} {\bibfnamefont {S.}~\bibnamefont {Miki}},
		\bibinfo {author} {\bibfnamefont {T.}~\bibnamefont {Yamashita}}, \bibinfo
		{author} {\bibfnamefont {H.}~\bibnamefont {Terai}}, \bibinfo {author}
		{\bibfnamefont {X.}~\bibnamefont {Jiang}}, \bibinfo {author} {\bibfnamefont
			{T.-Y.}\ \bibnamefont {Chen}}, \bibinfo {author} {\bibfnamefont
			{L.}~\bibnamefont {You}}, \bibinfo {author} {\bibfnamefont {X.}~\bibnamefont
			{Chen}}, \bibinfo {author} {\bibfnamefont {Z.}~\bibnamefont {Wang}}, \bibinfo
		{author} {\bibfnamefont {J.}~\bibnamefont {Fan}}, \bibinfo {author}
		{\bibfnamefont {Q.}~\bibnamefont {Zhang}}, \ and\ \bibinfo {author}
		{\bibfnamefont {J.}~\bibnamefont {Pan}},\ }\href {\doibase
		10.1038/nphoton.2016.179} {\bibfield  {journal} {\bibinfo  {journal} {Nature
				Photonics}\ }\textbf {\bibinfo {volume} {10}},\ \bibinfo {pages} {671}
		(\bibinfo {year} {2016})}\BibitemShut {NoStop}%
	\bibitem [{\citenamefont {Sun}\ \emph {et~al.}(2017)\citenamefont {Sun},
		\citenamefont {Jiang}, \citenamefont {Mao}, \citenamefont {You},
		\citenamefont {Zhang}, \citenamefont {Zhang}, \citenamefont {Jiang},
		\citenamefont {Chen}, \citenamefont {Li}, \citenamefont {Huang},
		\citenamefont {Chen}, \citenamefont {Wang}, \citenamefont {Fan},
		\citenamefont {Zhang},\ and\ \citenamefont
		{Pan}}]{sun2017_entswapping_100km_timebin}%
	\BibitemOpen
	\bibfield  {author} {\bibinfo {author} {\bibfnamefont {Q.-C.}\ \bibnamefont
			{Sun}}, \bibinfo {author} {\bibfnamefont {Y.-F.}\ \bibnamefont {Jiang}},
		\bibinfo {author} {\bibfnamefont {Y.-L.}\ \bibnamefont {Mao}}, \bibinfo
		{author} {\bibfnamefont {L.-X.}\ \bibnamefont {You}}, \bibinfo {author}
		{\bibfnamefont {W.}~\bibnamefont {Zhang}}, \bibinfo {author} {\bibfnamefont
			{W.-J.}\ \bibnamefont {Zhang}}, \bibinfo {author} {\bibfnamefont
			{X.}~\bibnamefont {Jiang}}, \bibinfo {author} {\bibfnamefont {T.-Y.}\
			\bibnamefont {Chen}}, \bibinfo {author} {\bibfnamefont {H.}~\bibnamefont
			{Li}}, \bibinfo {author} {\bibfnamefont {Y.-D.}\ \bibnamefont {Huang}},
		\bibinfo {author} {\bibfnamefont {X.-F.}\ \bibnamefont {Chen}}, \bibinfo
		{author} {\bibfnamefont {Z.}~\bibnamefont {Wang}}, \bibinfo {author}
		{\bibfnamefont {J.}~\bibnamefont {Fan}}, \bibinfo {author} {\bibfnamefont
			{Q.}~\bibnamefont {Zhang}}, \ and\ \bibinfo {author} {\bibfnamefont {J.-W.}\
			\bibnamefont {Pan}},\ }\href {\doibase 10.1364/OPTICA.4.001214} {\bibfield
		{journal} {\bibinfo  {journal} {Optica}\ }\textbf {\bibinfo {volume} {4}},\
		\bibinfo {pages} {1214} (\bibinfo {year} {2017})}\BibitemShut {NoStop}%
	\bibitem [{\citenamefont {Kim}\ \emph {et~al.}(2006)\citenamefont {Kim},
		\citenamefont {Fiorentino},\ and\ \citenamefont {Wong}}]{Kim2005}%
	\BibitemOpen
	\bibfield  {author} {\bibinfo {author} {\bibfnamefont {T.}~\bibnamefont
			{Kim}}, \bibinfo {author} {\bibfnamefont {M.}~\bibnamefont {Fiorentino}}, \
		and\ \bibinfo {author} {\bibfnamefont {F.~N.}\ \bibnamefont {Wong}},\ }\href
	{\doibase 10.1109/CLEO.2006.4628715} {\bibfield  {journal} {\bibinfo
			{journal} {Conference on Lasers and Electro-Optics and 2006 Quantum
				Electronics and Laser Science Conference, CLEO/QELS 2006}\ }\textbf {\bibinfo
			{volume} {73}},\ \bibinfo {pages} {12316} (\bibinfo {year}
		{2006})}\BibitemShut {NoStop}%
	\bibitem [{\citenamefont {Lim}\ \emph {et~al.}(2008)\citenamefont {Lim},
		\citenamefont {Yoshizawa}, \citenamefont {Tsuchida},\ and\ \citenamefont
		{Kikuchi}}]{lim2008}%
	\BibitemOpen
	\bibfield  {author} {\bibinfo {author} {\bibfnamefont {H.~C.}\ \bibnamefont
			{Lim}}, \bibinfo {author} {\bibfnamefont {A.}~\bibnamefont {Yoshizawa}},
		\bibinfo {author} {\bibfnamefont {H.}~\bibnamefont {Tsuchida}}, \ and\
		\bibinfo {author} {\bibfnamefont {K.}~\bibnamefont {Kikuchi}},\ }\href
	{\doibase 10.1364/OE.16.016052} {\bibfield  {journal} {\bibinfo  {journal}
			{Optics Express}\ }\textbf {\bibinfo {volume} {16}},\ \bibinfo {pages}
		{16052} (\bibinfo {year} {2008})}\BibitemShut {NoStop}%
	\bibitem [{\citenamefont {Clauser}\ \emph {et~al.}(1969)\citenamefont
		{Clauser}, \citenamefont {Horne}, \citenamefont {Shimony},\ and\
		\citenamefont {Holt}}]{CHSH}%
	\BibitemOpen
	\bibfield  {author} {\bibinfo {author} {\bibfnamefont {J.~F.}\ \bibnamefont
			{Clauser}}, \bibinfo {author} {\bibfnamefont {M.~A.}\ \bibnamefont {Horne}},
		\bibinfo {author} {\bibfnamefont {A.}~\bibnamefont {Shimony}}, \ and\
		\bibinfo {author} {\bibfnamefont {R.~A.}\ \bibnamefont {Holt}},\ }\href
	{\doibase 10.1103/PhysRevLett.23.880} {\bibfield  {journal} {\bibinfo
			{journal} {Physical Review Letters}\ }\textbf {\bibinfo {volume} {23}},\
		\bibinfo {pages} {880} (\bibinfo {year} {1969})}\BibitemShut {NoStop}%
	\bibitem [{\citenamefont {Cirel'son}(1980)}]{tsirelson1993quantum}%
	\BibitemOpen
	\bibfield  {author} {\bibinfo {author} {\bibfnamefont {B.~S.}\ \bibnamefont
			{Cirel'son}},\ }\href {\doibase 10.1007/BF00417500} {\bibfield  {journal}
		{\bibinfo  {journal} {Letters in Mathematical Physics}\ }\textbf {\bibinfo
			{volume} {4}},\ \bibinfo {pages} {93} (\bibinfo {year} {1980})}\BibitemShut
	{NoStop}%
	\bibitem [{\citenamefont {Ma}\ \emph {et~al.}(2007)\citenamefont {Ma},
		\citenamefont {Fung},\ and\ \citenamefont {Lo}}]{Ma2007}%
	\BibitemOpen
	\bibfield  {author} {\bibinfo {author} {\bibfnamefont {X.}~\bibnamefont
			{Ma}}, \bibinfo {author} {\bibfnamefont {C.-H.~F.}\ \bibnamefont {Fung}}, \
		and\ \bibinfo {author} {\bibfnamefont {H.-K.}\ \bibnamefont {Lo}},\ }\href
	{\doibase 10.1103/PhysRevA.76.012307} {\bibfield  {journal} {\bibinfo
			{journal} {Physical Review A}\ }\textbf {\bibinfo {volume} {76}},\ \bibinfo
		{pages} {012307} (\bibinfo {year} {2007})}\BibitemShut {NoStop}%
	\bibitem [{\citenamefont {Waddy}\ \emph {et~al.}(2005)\citenamefont {Waddy},
		\citenamefont {Chen},\ and\ \citenamefont {Bao}}]{waddy2005polarization}%
	\BibitemOpen
	\bibfield  {author} {\bibinfo {author} {\bibfnamefont {D.~S.}\ \bibnamefont
			{Waddy}}, \bibinfo {author} {\bibfnamefont {L.}~\bibnamefont {Chen}}, \ and\
		\bibinfo {author} {\bibfnamefont {X.}~\bibnamefont {Bao}},\ }\href {\doibase
		10.1016/j.yofte.2004.07.002} {\bibfield  {journal} {\bibinfo  {journal}
			{Optical Fiber Technology}\ }\textbf {\bibinfo {volume} {11}},\ \bibinfo
		{pages} {1} (\bibinfo {year} {2005})}\BibitemShut {NoStop}%
	\bibitem [{\citenamefont {Woodward}\ \emph {et~al.}(2014)\citenamefont
		{Woodward}, \citenamefont {Nelson}, \citenamefont {Schneider}, \citenamefont
		{Knox}, \citenamefont {O'Sullivan}, \citenamefont {Laperle}, \citenamefont
		{Moyer},\ and\ \citenamefont {Foo}}]{woodward2014long}%
	\BibitemOpen
	\bibfield  {author} {\bibinfo {author} {\bibfnamefont {S.~L.}\ \bibnamefont
			{Woodward}}, \bibinfo {author} {\bibfnamefont {L.~E.}\ \bibnamefont
			{Nelson}}, \bibinfo {author} {\bibfnamefont {C.~R.}\ \bibnamefont
			{Schneider}}, \bibinfo {author} {\bibfnamefont {L.~A.}\ \bibnamefont {Knox}},
		\bibinfo {author} {\bibfnamefont {M.}~\bibnamefont {O'Sullivan}}, \bibinfo
		{author} {\bibfnamefont {C.}~\bibnamefont {Laperle}}, \bibinfo {author}
		{\bibfnamefont {M.}~\bibnamefont {Moyer}}, \ and\ \bibinfo {author}
		{\bibfnamefont {S.}~\bibnamefont {Foo}},\ }\href {\doibase
		10.1109/LPT.2013.2290473} {\bibfield  {journal} {\bibinfo  {journal} {IEEE
				Photonics Technology Letters}\ }\textbf {\bibinfo {volume} {26}},\ \bibinfo
		{pages} {213} (\bibinfo {year} {2014})}\BibitemShut {NoStop}%
	\bibitem [{\citenamefont {Ding}\ \emph {et~al.}(2017)\citenamefont {Ding},
		\citenamefont {Chen}, \citenamefont {Wang}, \citenamefont {He}, \citenamefont
		{Yin}, \citenamefont {Chen}, \citenamefont {Zhou}, \citenamefont {Guo},\ and\
		\citenamefont {Han}}]{ding2017polarization}%
	\BibitemOpen
	\bibfield  {author} {\bibinfo {author} {\bibfnamefont {Y.-Y.}\ \bibnamefont
			{Ding}}, \bibinfo {author} {\bibfnamefont {H.}~\bibnamefont {Chen}}, \bibinfo
		{author} {\bibfnamefont {S.}~\bibnamefont {Wang}}, \bibinfo {author}
		{\bibfnamefont {D.-Y.}\ \bibnamefont {He}}, \bibinfo {author} {\bibfnamefont
			{Z.-Q.}\ \bibnamefont {Yin}}, \bibinfo {author} {\bibfnamefont
			{W.}~\bibnamefont {Chen}}, \bibinfo {author} {\bibfnamefont {Z.}~\bibnamefont
			{Zhou}}, \bibinfo {author} {\bibfnamefont {G.-C.}\ \bibnamefont {Guo}}, \
		and\ \bibinfo {author} {\bibfnamefont {Z.-F.}\ \bibnamefont {Han}},\ }\href
	{\doibase 10.1364/OE.25.027923} {\bibfield  {journal} {\bibinfo  {journal}
			{Optics Express}\ }\textbf {\bibinfo {volume} {25}},\ \bibinfo {pages}
		{27923} (\bibinfo {year} {2017})}\BibitemShut {NoStop}%
	\bibitem [{\citenamefont {Gisin}\ and\ \citenamefont
		{Thew}(2007)}]{gisin2007quantumcommunication}%
	\BibitemOpen
	\bibfield  {author} {\bibinfo {author} {\bibfnamefont {N.}~\bibnamefont
			{Gisin}}\ and\ \bibinfo {author} {\bibfnamefont {R.}~\bibnamefont {Thew}},\
	}\href {\doibase 10.1038/nphoton.2007.22} {\bibfield  {journal} {\bibinfo
			{journal} {Nature Photonics}\ }\textbf {\bibinfo {volume} {1}},\ \bibinfo
		{pages} {165} (\bibinfo {year} {2007})}\BibitemShut {NoStop}%
\end{thebibliography}
\end{document}